\documentclass[twocolumn,amsmath,amssymb,pra,superscriptaddress]{revtex4}

\usepackage{graphics}
\usepackage{epsfig}
\usepackage{amsmath}
\usepackage{amsfonts}
\usepackage{color}
\usepackage{dcolumn}
\usepackage{bm}
\usepackage{mathrsfs}

\newcommand{\ds}{\displaystyle}

\newcommand{\ket}[1]{| #1 \rangle}

\newcommand{\opij}[3]{\langle #1 | #2 | #3 \rangle}
\newcommand{\braket}[2]{\langle #1 | #2 \rangle}
\newcommand{\eci }{\mathcal{E}}

\def\oper#1{\hat{\rm #1}}

\begin{document}

  \bibliographystyle{apsrev}

\title{Theoretical description of two ultracold atoms in
         finite 3D optical lattices using realistic interatomic
         interaction potentials}

       \author{Sergey Grishkevich}
       \affiliation{Theoretische
           Physik, Universit\"at des Saarlandes, 66041 Saarbr\"ucken,
           Germany}

       \author{Simon Sala}

       \author{Alejandro Saenz}
       \affiliation{AG Moderne Optik, Institut f\"ur Physik,
         Humboldt-Universit\"at zu Berlin, Newtonstr.~15,
         12\,489 Berlin, Germany}

%   \title{Theoretical description of two ultracold atoms in 
%          finite 3D optical lattices using realistic interatomic 
%          interaction potentials}

%        \author{Sergey Grishkevich}
%        \affiliation{ Theoretische Physik, 
%                      Universit\"at des Saarlandes, D-66041 Saarbr\"ucken, Germany}
%        \author{Simon Sala}
%        \affiliation{AG Moderne Optik, Institut f\"ur Physik,
%          Humboldt-Universit\"at zu Berlin, Newtonstr.~15,
%          12\,489 Berlin, Germany}
%        \author{Alejandro Saenz}
%        \affiliation{AG Moderne Optik, Institut f\"ur Physik,
%          Humboldt-Universit\"at zu Berlin, Newtonstr.~15,
%          12\,489 Berlin, Germany}

 %      \date{\today}

  \begin{abstract}
     A theoretical approach is described for an exact numerical 
     treatment of a pair of ultracold atoms interacting via a central 
     potential that are trapped in a finite three-dimensional optical 
     lattice. The coupling of center-of-mass and relative-motion 
     coordinates is treated using an exact diagonalization  
     (configuration-interaction) approach. 
     The orthorhombic symmetry of an optical lattice with 
     three different but orthogonal lattice vectors is explicitly 
     considered as is the Fermionic or Bosonic symmetry in the case 
     of indistinguishable particles. 
  \end{abstract}

    \maketitle

\section{Introduction}
\label{sec:intro}
 The physics of ultracold quantum gases attracts a lot of interest since 
 the experimental observation of Bose-Einstein condensation in dilute 
 alkali-metal atom gases~\cite{cold:ande95,cold:davi95}. Besides the exciting 
 physics of ultracold gases by itself, a further important progress 
 was the loading of the ultracold gas into an optical lattice formed with
 the aid of standing light waves~\cite{cold:jaks98,cold:grei02,cold:koeh05}. 
 The optical lattice resembles in some sense the periodicity of a crystal
 potential~\cite{cold:meac98,cold:bloc05,cold:lewe07,cold:bloc08}, 
 but is practically 
 free of phonons. In contrast to real solids the lattice parameters are, 
 in addition, easily tunable by a variation of the laser intensity, 
 the relative orientation of the laser beams, or the wavelength. 
 In the first case the lattice depth, in the other cases the lattice 
 geometry can be manipulated.

 Moreover, in the ultracold regime the effective long-range
 interaction between atoms is usually well characterized by a single
 parameter, the scattering length~\cite{cold:wein99}. This simplifies the investigation of
 atomic systems in the ultracold regime. The effective interaction can be
 either attractive or repulsive, depending on the type of atoms involved. While different
 kinds of chemical elements, their isotopes, or atoms in different electronic 
 or spin states cover already a wide range of interaction strengths, 
 an almost full tunability of the atom-atom interactions in ultracold gases 
 is achieved using magnetic Feshbach 
 resonances~\cite{cold:loft02,cold:rega03}. Close to the
 resonance value of the magnetic field the scattering length diverges and 
 the effective interaction varies in a wide range, in principle from 
 being infinitely strongly repulsive to infinitely strongly attractive. 
 This possibility of active control of the interparticle interaction 
 makes ultracold atoms in optical lattices an ideal tool for, e.\,g., 
 exploring the properties of many-body Hamiltonians describing particles in 
 periodic potentials, like the Hubbard model~\cite{cold:hubb63,cold:jaks98}. 
 Examples are the experimental studies of a Bosonic Mott
 insulator~\cite{cold:grei02}, a Fermionic band
 insulator~\cite{cold:koeh05}, or the simulation of antiferromagnetic 
 spin chains in an optical lattice~\cite{cold:simo11}.

 A further important aspect of ultracold quantum gases in optical lattices is
 the in principle arbitrary filling that can be realized, while the filling is
 strongly constrained in usual solid-state systems by charge
 neutrality. Ultracold quantum gases thus allow studies far away from the
 often considered half-filling case. One interesting limit is the one of very
 sparsely populated lattices in which few-body quantum dynamics can be studied
 very accurately~\cite{cold:krae06}. This initiated recently a number
 of corresponding theoretical studies~\cite{cold:vali10,cold:vali10b,cold:chat10}.
 These investigations are often further motivated by the fact that the
 direct experimental control of many parameters
 together with the high degree of coherence that is not destroyed by
 phonons
 led to proposals to use ultracold quantum gases in optical lattices in
 quantum-information applications like quantum simulators or even quantum
 computers~\cite{cold:demi02,cold:jaks04,cold:lewe07,cold:bloc08,cold:simo11,cold:schn11a}. For such applications a very precise knowledge
 about the microscopic interactions between quantum particles in an optical
 lattice is a prerequisite.

 At the required level of accuracy a description of the atoms as simply 
 being trapped in an array of harmonic potentials becomes inappropriate. 
 For example, during a controlled quantum-gate operation it is usually 
 necessary to bring two atoms from different sites into contact with each 
 other. Atoms in different electronic states may be involved in such an 
 operation, as those states may encode the two qubit states ($\ket{0}$ 
 and $\ket{1}$). However, in such a case the center-of-mass (c.m.) and 
 relative (rel.) motions of two atoms even in the same potential well do 
 not separate, even not within the harmonic 
 approximation~\cite{cold:bold05,cold:gris07}. This is due 
 to the fact that different hyperfine states are usually accompanied 
 by different polarizabilites. Thus the atoms in different states 
 experience different trapping frequencies even within the harmonic
 approximation. Evidently, such a non-separability of c.m.\ and
 rel.\ motions always occurs for heteronuclear systems (different 
 atomic species) with different masses and polarizabilites. 
 Experimental evidence for the corresponding breakdown of the harmonic 
 approximation for a heteronuclear system was given in \cite{cold:ospe06a} 
 and theoretically confirmed \cite{cold:deur08,cold:gris09}.   
 If the considered atoms are not tightly bound in the same potential 
 well and thus if the multi-well structure of the optical lattice is 
 important, there is evidently no separability of c.m.\ and rel.\ motion, 
 even not for identical atoms. 

 Already the theoretical treatment of two atoms in an optical lattice is a
 formidable task, if realistic atom-atom interaction needs to be
 considered. While this interaction may often be described by a central force,
 the interaction potential stems from laborious quantum-chemistry calculations
 and is thus only numerically given.  The transition to c.m.\ and rel.\
 coordinates simplifies the problem dramatically, since the interatomic
 interaction affects only the rel.\ motion and in the case of an isotropic
 interaction even only the radial part of it. However, the above-mentioned
 non-separability demands finally to treat the full six-dimensional problem.
 Furthermore, the matrix elements describing the interaction with the trapping
 potential become more involved, since they do not separate in c.m.\ and rel.\
 coordinates. Alternatively, the problem may be solved in (absolute) Cartesian
 coordinates. In this case the potential describing the optical lattice
 separates for both particles and, e.\,g., in the orthorhombic case even for
 the three Cartesian coordinates.  However, in this case the particle
 interaction terms restore the non-separability of the six-dimensional
 problem, except for very special cases like the $r^{2n}$ potentials 
 \cite{cold:arms11}  where $r$ is the radial coordinate of
 rel.\ motion and $n$ is an integer.  In view of its universality with respect
 to the interparticle interaction the treatment in c.m.\ and rel.\ coordinates
 is favorable.  Furthermore, as will be shown in detail below, the use of
 Taylor expansions of the optical-lattice potential allows for a reasonable
 efficiency.

 In this work a numerical approach is presented for the theoretical 
 treatment of two particles that interact {\it via} a central 
 (isotropic) interaction potential and are trapped in a finite 
 orthorhombic $\sin^2$- or $\cos^2$-type periodic potential. 
 While the main motivation is the treatment of ultracold atoms and 
 molecules in optical lattices, the code allows also to treat 
 other particles and was, e.\,g., recently also employed in a 
 study of electrons and excitons in quantum-dot
 molecules.  
 The uncoupled Schr\"odinger equations for c.m.\ and rel.\ motion 
 are solved by an expansion of the radial parts in $B$ splines 
 and the angular parts in spherical harmonics. The coupling 
 is then considered by means of configuration interaction 
 (exact diagonalization). The orthorhombic D$_{\rm 2h}$ 
 symmetry is fully accounted for. 
 
 The approach was already 
 successfully applied in a systematic investigation of the effects 
 of anharmonicity and coupling of c.m.\ and rel.\ motion 
 for two atoms in a single well of an optical lattice \cite{cold:gris09}. 
 Together with the experimental results in \cite{cold:ospe06a} 
 this allowed the conclusion that only the inclusion of the effects 
 of anharmonicity and coupling (and thus deviations from a
 simple uncoupled harmonic model) lead to agreement with experiment. 
 Furthermore, considering a triple-well potential the optimal 
 Bose-Hubbard parameter were obtained and the range of validity 
 of the Bose-Hubbard model was explored quantitatively \cite{cold:schn09}. 
 Such a study is of importance, as it provides a link to many-body physics 
 and large lattices within the most popular model for the description 
  of ultracold atoms in optical lattices \cite{cold:jaks98}. 
 It should be emphasized that in view of proposed quantum-information 
 applications the physics of, e.\,g., few atoms in double-well 
 potentials is, however, already of interest by itself 
 \cite{cold:milb97,cold:ande06,cold:foel07,cold:chei08,
 cold:sebb06,cold:trot08}. The triple-well system was on the other hand,
 e.\,g., proposed to serve as a transistor, where the population of the 
 middle well controls the tunneling of particles from the left to the right 
 well \cite{cold:stic07}.  
 
 The paper is organized in the following way. In Sec.~\ref{sec:system} 
 the system and its Hamiltonian is introduced. This includes 
 the choice of coordinate system in Sec.~\ref{subsec:hamilt}, 
 the Taylor expansion of the optical-lattice potential in 
 Sec.~\ref{subsec:taylor}, and the consequent expansion of its 
 angular part in spherical harmonics in Sec.~\ref{subsec:spherical}. 
 The obtained final form of the Hamiltonian and the alternative 
 $\cos^2$ lattice are discussed in Secs.~\ref{subsubsec:finalH} and 
 \ref{subsubsec:finite}, respectively.  
 Sec.~\ref{sec:hamandwf} describes the exact diagonalization approach 
 with the corresponding Schr\"odinger equations in 
 Sec.~\ref{subsec:schroedinger} and all matrix elements that have to 
 be calculated in Sec.~\ref{subsec:matel}.
 The implementation of symmetry into the approach is described 
 in Sec.~\ref{sec:symofsys}. 
 In Sec.~\ref{sec:compdet} computational details are given. This includes 
 practical aspects of the interaction potential in 
 Sec.~\ref{subsec:interaction}, basis-set considerations in 
 Sec.~\ref{subsec:basisset}, an example calculation of rel.\ motion 
 orbitals for a highly anisotropic trap potential including a 
 convergence study in Sec.~\ref{subsec:example}, and practical 
 issues of the final exact-diagonalization step in 
 Sec.~\ref{subsec:exdiag}. The paper closes with a brief summary and 
 outlook in Sec.~\ref{sec:outlook}.

 \section{Hamiltonian for two atoms in an optical lattice}
 \label{sec:system}

 \subsection{System and coordinates}
 \label{subsec:hamilt}

 The Hamiltonian describing two interacting atoms with 
 coordinate vectors $\vec{r}_1$, $\vec{r}_2$ that are trapped in a
 three-dimensional optical lattice is given by 
\begin{equation}
     \oper{\mathcal{H}}(\vec{r}_1,\vec{r}_2)
           =
           \sum_{j=1}^2
%	   \left[
             \oper{\mathcal{H}}_j(\vec{r}_j)
%              \oper{\mathcal{T}}(\vec{r}_j) +
% 		\oper{\mathcal{V}}(\vec{r}_j)
%	   \right]
           + \oper{\mathcal{U}}(\vec{r}_1,\vec{r}_2)
   \label{eq:origham}
 \end{equation}
with
\begin{equation}
  \label{eq:single_part_hamil}
  \oper{\mathcal{H}}_j(\vec{r}_j)= \oper{\mathcal{T}}_j(\vec{r}_j) +
  \oper{\mathcal{V}}_j(\vec{r}_j) 
\end{equation}
 where  $\oper{\mathcal{T}}_j$ is the
 one-particle kinetic energy operator, $\oper{\mathcal{V}}_j$ is the trapping
 potential of the optical lattice for particle $j$, and $\oper{\mathcal{U}}$
 is the atom-atom interaction potential.  If the lattice is formed by three
 counterpropagating laser fields that are orthogonal to each other, the atoms
 experience the periodic potential
\begin{equation}
   \oper{\mathcal{V}}_j(\vec{r}_j) =
           \sum\limits_{c={x,y,z}}
           V_c^j \: \sin^2(k_c c_j) \, ,
  \label{eq:sinform}
 \end{equation}
 due to the dipole forces, if the laser frequencies are sufficiently 
 far-detuned from resonant transitions. In Eq.~(\ref{eq:sinform}) 
 $V_c^j$ is the potential depth acting on particle $j$ along the 
 direction $c$ ($=x, y, z$) and is equal to the product of the
 laser intensity $I_c$ and the polarizability of the particle $j$. 
 Furthermore, $k_c=2\pi/\lambda_c$ is the wave vector and 
 $\lambda_c$ is the wavelength of the laser that creates the lattice 
 potential along the coordinate $c$. 

\begin{figure}[!ht]
  \centering
  \includegraphics[width=0.32\textwidth]{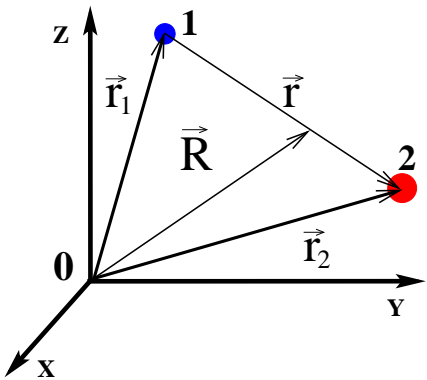}
  \caption[The absolute, c.m.-rel.\ and spherical coordinates]
      {{\footnotesize
          Two particles 1 and 2 in the absolute and c.m.-rel.\
          Cartesian coordinate systems.
    } }
  \label{fig:coord}
 \end{figure}

 A direct solution of the Schr\"odinger equation with the Hamiltonian
 given in the form of Eq.\,(\ref{eq:origham}) is complicated, since
 $\oper{\mathcal{U}}$ depends in general on all six coordinates describing the
 two-particle system, even if the atom-atom interaction is central,
 i.\,e., $\oper{\mathcal{U}}(\vec{r}_1,\vec{r}_2)=
 \oper{\mathcal{U}}(|\vec{r}_1 - \vec{r}_2|) =: \oper{u}(r)$ with 
 $r=|\vec{r}_1 - \vec{r}_2|$. For
 realistic interatomic interaction potentials, there is no separability
 and this leads to very demanding six-dimensional integrals. Therefore,
 it is more convenient to treat the two-particle problem in the
 c.m.\ and rel.\ coordinates, $\vec{R}$ and $\vec{r}$ respectively, defined as
  \begin{equation}
   \ds 
   \vec{r} = \vec{r}_1 - \vec{r}_2 \, ,
   \quad
   \vec{R} = \mu_1 \vec{r}_1 + \mu_2 \vec{r}_2 \, 
   \label{eq:toabs}
 \end{equation}
 with the dimensionless parameters  $\ds \mu_1 = m_1/(m_1+m_2)$ and
 $\ds \mu_2 = m_2/(m_1+m_2)$ where
 $m_j$ is the mass of the $j$th particle. The system of two
 atoms in a 3D space as well as the c.m.-rel.\ coordinate system is
 sketched in Fig.~\ref{fig:coord}.  The evident advantage of 
 this choice of coordinates is the fact that 
 the interaction potential acts only 
 on the rel.\ coordinate $\vec{r}$ and thus on three instead of six 
 dimensions. If spherical
 coordinates are adopted, a central interaction potential  
 $\oper{u}(r)$
 depends even on the radial coordinate $r$ only.

 On the other hand, the formulation of the two-particle problem in
 the c.m.\ and rel.\ coordinates complicates the treatment of the
 trapping potential, because its original separability in the absolute 
 Cartesian coordinates $\vec{r}_j$ is lost in the c.m.-rel.\ system. 
 Only within the harmonic approximation for the trapping potential 
 and for two identical atoms in the same internal state there is 
 complete separability in c.m.-rel.\ coordinates \cite{cold:blum02}.  
 If the true atom-atom interaction is furthermore replaced by 
 a $\delta$-function pseudopotential, the corresponding Schr\"odinger 
 equation possesses an analytical solution for isotropic and 
 some anisotropic harmonic traps~\cite{cold:busc98,cold:idzi05}. 
 However, even within the harmonic approximation the separability is 
 lost, if the two atoms experience different trapping potentials. 
 This is the case, if a heteronuclear system or two identical atoms 
 in different electronic states are considered. In the general case 
 of a treatment beyond the harmonic approximation and, especially,   
 if the atoms are spread over more than one potential well, there 
 is evidently no separability at all. 

 In order to keep the flexibility with respect to the interparticle 
 interaction and the advantage of its simple handling by using 
 spherical c.m.\ and rel.\ coordinates, the optical-lattice potential 
 has to be brought into a form that is convenient for its numerical 
 treatment in those coordinates. This is done in two steps. First, 
 a Taylor expansion of the sinusoidal trapping potential 
 (\ref{eq:sinform}) is performed in Cartesian c.m.\ and rel.\ coordinates 
 (Sec.~\ref{subsec:taylor}). The result is then 
 transformed into spherical coordinates using an expansion in spherical 
 harmonics (Sec.~\ref{subsec:spherical}). 

 \subsection{Taylor expansion of the optical-lattice potential}
 \label{subsec:taylor}

 The optical-lattice potential for both particles in 
 Eq.~(\ref{eq:sinform}) is given in the Cartesian c.m.-rel.\
 coordinates $\vec{R}_c$ and $\vec{r}_c$ ($c=x,y,z$) as 
 \begin{align}
     \ds 
    \oper{{V}} (\vec{R},\vec{r}\,) &= 
     \sum\limits_{i=1}^2 \sum\limits_{c=x,y,z}
     V_c^i 
 \nonumber \\
     &\hspace{1cm}\times
     \sin^2(k_c [R_c + (-1)^{i-1}\mu_{\eta_i} r_c])
  \label{eq:olfreepar}
 \end{align}
 where the index $\ds \eta_i = i+(-1)^{i-1}$ was introduced for 
 compactness. Using the standard trigonometric relations the 
 optical-lattice potential can be rewritten in the more suitable form  
   \begin{multline}
     \ds 
         \oper{V}(\vec{R},\vec{r}\,) =
         \frac12 \sum\limits_{i=1}^2
         \sum\limits_{c=x,y,z}
         V^i_c 
         \left[1 + (-1)^{\eta_i} \sin (2k_c R_c) \right.
         {}
          \\
         {}
         \left.
         \times\,
         \sin (2k_c r_c \mu_{\eta_i})
         -
         \cos (2k_c R_c)
         \cos (2k_c r_c \mu_{\eta_i})
         \right] 
         \, .
   \label{eq:olcomreltrigrel}
   \end{multline}
 In order to achieve a maximal separation of the coordinates 
 $\vec{R}$ and $\vec{r}$ the trigonometric functions in 
 Eq.~(\ref{eq:olcomreltrigrel}) are expanded in Taylor series 
 around the origin of the c.m.\ and rel.\ coordinates,
 \begin{gather}
   \ds
     \sin (2k_c R_c)
     \sin (2k_c r_c \mu_{\eta_s}) = 
     \sum\limits_{j=0}^{\infty}
     \sum\limits_{i=0}^{\infty}
     \frac{(-1)^{i+j}}{(2i+1)!(2j+1)!}
     \nonumber \\
     \times\, (2k_c)^{2i+1} (2k_c\mu_{\eta_s})^{2j+1}
     R_c^{2i+1} r_c^{2j+1} \, ,
  \label{eq:trigprodexpinfsin}
 \end{gather}
 \begin{gather}
       \cos (2k_c R_c)
       \cos (2k_c r_c \mu_{\eta_s}) = 
       \sum\limits_{t=0}^{\infty}
       \sum\limits_{k=0}^{\infty}
       \frac{(-1)^{k+t}}{(2k)!(2t)!}
       \nonumber \\
       \times\, (2k_c)^{2k} (2k_c\mu_{\eta_s})^{2t}
       R_c^{2k} r_c^{2t} \, .
  \label{eq:trigprodexpinf}
  \end{gather}
 In a practical numerical implementation, the infinite sum must be
 truncated. If, for example, the expansion is restricted up to the
 $\ds (2n)$th degree (with the order $n=1,2,3, ...$), the infinite
 summations in Eqs.~(\ref{eq:trigprodexpinfsin})
 and~(\ref{eq:trigprodexpinf}) are changed, since the indices 
 fulfill 
 \begin{align}
   \ds & 2i+1+2j+1\le 2n \quad \text{with}
       \nonumber \\
       & i\le n-1-j \quad 
       \text{and} \quad j\le n-1\, ,
        \smallskip 
  \nonumber \\
       & 2k+2t \le 2n \quad \text{with}
       \nonumber \\
       & k\le n-t  \quad 
       \text{and} \quad t\le n \, .
   \label{eq:sumlim}
 \end{align}
 Hence, the optical-lattice potential can be approximated by 
 a Taylor expansion of degree $\ds (2n)$ as
 \begin{multline}
    \ds 
        \oper{V}(\vec{R},\vec{r}) \approx 
        \frac12 \sum\limits_{s=1}^2 
        \sum\limits_{c=x,y,z}
        V^s_c \Biggl[1 + (-1)^{\eta_s}
        \sum\limits_{j=0}^{n-1}
          {}
         \\
         {}
        \times\,
        \sum\limits_{i=0}^{n-1-j} 
        \mathbb{C}_{ijcs}^{\text{sin}} 
        R_c^{2i+1} r_c^{2j+1}
%          \nonumber \\
         -\,
        \sum\limits_{t=0}^{n}
        \sum\limits_{k=0}^{n-t} 
        \mathbb{C}_{tkcs}^{\text{cos}} 
        R_c^{2k} r_c^{2t} \Biggr]
  \label{eq:olexpandfin}
 \end{multline}
 where the coefficients
 \begin{equation}
   \ds \mathbb{C}_{tkcs}^{\text{cos}} =
    \frac{(-1)^{k+t}}{(2k)!(2t)!}
    (2k_c)^{2(k+t)} \mu_{\eta_s}^{2t}
    \, ,
  \end{equation}
 \begin{equation}
    \ds \mathbb{C}_{ijcs}^{\text{sin}} =
    \frac{(-1)^{i+j}}{(2i+1)!(2j+1)!}
    (2k_c)^{2(i+j+1)} \mu_{\eta_s}^{2j+1}
 \end{equation}
 are introduced for compactness.

 Using Eq.~(\ref{eq:olexpandfin}) it is possible to split 
 the optical-lattice potential according to
 \begin{equation}
 \oper{V}(\vec{R},\vec{r}) = 
     \oper{v}_{\rm c.m.} (\vec{R}) + \oper{v}_{\rm rel.} (\vec{r}) 
              +   \oper{W} (\vec{R},\vec{r})     
 \label{eq:splitting}
 \end{equation}
 into $\oper{v}_{\rm c.m.}$ and $\oper{v}_{\rm rel.}$ that contain all 
 terms depending solely on the c.m.\ coordinate and the rel.\ 
 coordinate, respectively. The coupling terms between c.m.\ and 
 rel.\ motion are now contained in $\oper{W}$. The three 
 components of $\oper{{V}}$ are
  \begin{equation}
    \ds
         \oper{v}_{\rm c.m.}
             (\vec{R}) = 
             -\frac12 \sum\limits_{s=1}^2 
             \sum\limits_{c=x,y,z}
             V^s_c \sum\limits_{k=1}^{n}
             \mathbb{C}_{0kcs}^{\text{cos}} 
             R_c^{2k} 
             \label{eq:expcom}
  \end{equation}
  \begin{equation}
    \ds
         \oper{v}_{\rm rel.}
             (\vec{r}) =
             -\frac12 \sum\limits_{s=1}^2
             \sum\limits_{c=x,y,z}
             V^s_c \sum\limits_{t=1}^{n}
             \mathbb{C}_{t0cs}^{\text{cos}}
             r_c^{2t}
             \label{eq:exprel}
   \end{equation}
  \begin{multline}
     \ds
         \oper{W}
             (\vec{R},\vec{r}) =
         \frac12 \sum\limits_{s=1}^2
             \sum\limits_{c=x,y,z}
             V^s_c \Biggl[ (-1)^{\eta_s}
             {}
              \\
             {}
             \times\, \sum\limits_{j=0}^{n-1}
             \sum\limits_{i=0}^{n-1-j}
             \mathbb{C}_{ijcs}^{\text{sin}}
             R_c^{2i+1} r_c^{2j+1} -  \sum\limits_{t=1}^{n}
             \sum\limits_{k=1}^{n-t}
             \mathbb{C}_{tkcs}^{\text{cos}}
             R_c^{2k} r_c^{2t} \Biggr] \, .
             \label{eq:expcoupl}
 \end{multline}
 Note that the sum $\sum\limits_{k=1}^{0}$ that occurs for $n=1$ in 
 the second term of Eq.~(\ref{eq:expcoupl}) does not indicate an inverse 
 summation but its absence, i.e., no sum at all.

 As a result of the Taylor expansion the Hamiltonian (\ref{eq:origham}) 
 is transformed into the more convenient form 
 \begin{equation}
   \label{eq:fullham}
   \oper{H}(\vec{R},\vec{r}\,) = \oper{h}_{\rm c.m.}(\vec{R}\,) 
                              + \oper{h}_{\rm rel.}(\vec{r}\,)
                              + \oper{W}(\vec{R},\vec{r}\,) 
 \end{equation}
  with
 \begin{gather}
      \ds \oper{h}_{\rm c.m.}(\vec{R}\,) = \oper{t}_{\rm c.m.} (\vec{R}\,)
                                + \oper{v}_{\rm c.m.}(\vec{R}\,),
                                \label{eq:hamcm}
           \\ 
	   \nonumber \\
          \ds \oper{h}_{\rm rel.}(\vec{r}\,) = \oper{t}_{\rm rel.}(\vec{r}\,)
                                + \oper{v}_{\rm rel.}(\vec{r}\,)
                                + \oper{u} (r)
                                \label{eq:hamrm}  
 \end{gather}
 where we introduced $\oper{t}_{\rm rel.}$ and $\oper{t}_{\rm c.m.}$ 
 for the kinetic-energy 
 operators of the c.m.\ and the rel.\ motion, respectively. 
 It is worth emphasizing that in the present formulation only the
 truly non-separable terms (represented by products of the c.m.\ and
 rel.\ coordinates) are left in the coupling term $\oper{W}$.
 All separable terms of the optical lattice potential are included into
 the c.m.\ and rel.\ Hamiltonians $\oper{h}_{\rm c.m.}$ and
 $\oper{h}_{\rm rel.}$, respectively.

 In fact, there is a specific case in which the optical-lattice 
 potential in Eq.~(\ref{eq:olcomreltrigrel}) can be brought 
 into the form of Eq.~(\ref{eq:splitting}) and thus the Hamiltonian 
 (\ref{eq:fullham}) can be obtained without performing 
 the Taylor expansion. This is the case for two identical 
 particles that are both in the same state, if they are deposited 
 in a cubic lattice with equal intensities and $k$ vectors along 
 each of the spatial directions \cite{cold:bold05}.
 If all these conditions are satisfied, then
 Eq.~(\ref{eq:olcomreltrigrel}) can be written as
 \begin{gather}
     \ds
         \oper{v}_{\rm c.m.}(\vec{R}) =
            2 V_0 \sum\limits_{c=x,y,z}
            \sin^2 \left(k R_c \right) \,,
     \\ \ds
         \oper{v}_{\rm rel.}(\vec{r}) =
           2 V_0 \sum\limits_{c=x,y,z}
           \sin^2 \left(\frac{k r_c}2\right) \,,
     \\ \ds
         \oper{W}(\vec{R},\vec{r}) =
           -4 V_0 \sum\limits_{c=x,y,z}
           \sin^2 (k R_c ) \sin^2 \left(\frac{k r_c}2\right)\,.
 \end{gather}

 Noteworthy, the sum of the $\sin^2$-shaped lattice potentials for 
 the two individual particles transforms into $\sin^2$-shaped 
 potentials for both the 
 c.m.\ and rel.\ motion, though the one of the c.m.\ motion possesses 
 a different periodicity. In fact, this is also true in the here 
 considered general case of a heteronuclear atom pair in an orthorhombic 
 lattice. This is easily seen by extending the Taylor expansions  
 in Eqs.~(\ref{eq:expcom}) to (\ref{eq:expcoupl}) which gives   
  \begin{equation}
    \ds
         \lim_{n\rightarrow\infty} 
             \oper{v}_{\rm c.m.}
             (\vec{R}) = 
             \sum\limits_{s=1}^2 
             \sum\limits_{c=x,y,z}
             V^s_c \sin^2 (k_c R_c) \, ,
             \label{eq:expcom2}
  \end{equation}
  \begin{equation}
    \ds
         \lim_{n\rightarrow\infty} 
         \oper{v}_{\rm rel.}
             (\vec{r}) =
             \sum\limits_{s=1}^2
             \sum\limits_{c=x,y,z}
             V^s_c \sin^2 (k_c r_c
             \mu_{\eta_s}) \, ,
             \label{eq:exprel2}
   \end{equation}
  \begin{multline}
     \ds
         \lim_{n\rightarrow\infty} 
         \oper{W}
             (\vec{R},\vec{r}) =
         \frac12 \sum\limits_{s=1}^2
             \sum\limits_{c=x,y,z}
             V^s_c \Biggl[ (-1)^{\eta_s}
             {}
              \\
             {}
             \sin (2k_c R_c)\sin (2k_c 
             r_c\mu_{\eta_s})
             {}
              \\
             {}
           -\, \sum\limits_{t=1}^{n}
             \sum\limits_{k=1}^{n-t} 
             \mathbb{C}_{tkcs}^{\text{cos}} 
             R_c^{2k} r_c^{2t} \Biggr] \, .
             \label{eq:expcoupl2}
 \end{multline}

 Since an analytical solution for the $\sin^2$-like lattice and
 non-interacting particles exists \cite{cold:kohn59}, those 
 uncoupled known solutions of the Schr\"odinger equations 
 with the Hamiltonians in Eqs.~(\ref{eq:expcom2}) and (\ref{eq:exprel2}) 
 could be used as a basis for solving the coupled problem. However, 
 due to the presence of the central interaction potential a 
 transformation to the spherical coordinate system is advantageous, 
 since
 $\oper{\mathcal{U}}(\vec{r})=\oper{\mathcal{U}}(\sqrt{x^2+y^2+z^2})$
 does normally not allow for 
 a simple solution in Cartesian coordinates. However, this change 
 of coordinates is inconvenient, since the 
 mentioned analytical solutions of the $\sin^2$ potential do not 
 split into simple products of the radial and angular parts. 
 Since $\oper{\mathcal{U}}$ can in principle have any possible
 functional form, 
 it is more convenient to transform the optical-lattice potential 
 into a form that is suitable for a calculation in spherical 
 coordinates. This is done in the following subsection by an 
 expansion in spherical harmonics.

 \subsection{Expansion of the optical-lattice potential in spherical 
harmonics}
 \label{subsec:spherical}

 \subsubsection{Auxiliary functions $\mathbb{Y}_{lmt}^c$ and 
          $\tilde{\mathbb{Y}}_{lmt}^c$}
 \label{subsubsec:auxiliary}

 In order to express the optical-lattices potentials  
 $\oper{v}_{\rm c.m.}$, $\oper{v}_{\rm rel.}$, and $\oper{W}$ in 
 Eqs.~(\ref{eq:expcom}), (\ref{eq:exprel}), and~(\ref{eq:expcoupl}),
 respectively, in terms of the spherical harmonics $Y_l^m$, the 
 corresponding polynomials in the Cartesian coordinates $\vec{r}_c^{\,t}$ 
 and $\vec{R}_c^t$ have to be rewritten as radial part times an 
 angular function, i.\,e., as $r_c^t\,F_t^c(\theta,\phi)$ and 
 $R_c^t\,F_t^c(\theta,\phi)$, respectively. Every function 
 of the angles $\phi$ and $\theta$ and thus $F_t^c$ can be expanded in 
 spherical harmonics according to    
 \begin{equation}
   \ds F_t^c(\theta,\phi)=
    \sum\limits_{l=0}^{\infty}
    \sum\limits_{m=-l}^{l}
    \mathbb{Y}_{lmt}^c Y_l^m
    (\theta,\phi) 
    \label{eq:sphharexp}
 \end{equation}
 where the projection coefficients $\mathbb{Y}^c_{lmt}$ are 
 given as
 \begin{gather}
   \ds \mathbb{Y}_{lmt}^c =
   (-1)^{m} A_{l-m}
   \int_\Omega d\Omega\, F_t^c(\theta,\phi)
    {P_{l}^{-m}}(\theta) e^{-{\rm i }m\phi}\, .
   \label{eq:sphercoef}
 \end{gather}
 In Eq.~(\ref{eq:sphercoef}) ${\rm i}$ stands for the imaginary unit,
 $\ds P_l^m(\theta)$ are the associated Legendre polynomials and 
 $\ds A_{l-m}$
 is a constant prefactor which is defined by %
 \begin{equation}
   \ds A_{lm}=\sqrt{\frac{2l+1}{4\pi}\frac{(l-m)!}{(l+m)!}} \, .
 \end{equation}
 Finally, the integral over the angular arguments is
 $\int_\Omega d\Omega =\int\limits_{0}^{\pi} d \theta
 \int\limits_{0}^{2 \pi} d \phi\sin(\theta)$. Due to their different 
 properties, it is useful to distinguish two types of the expansion
 coefficients $\mathbb{Y}_{lmt}^c$, those with even values of $t$ 
 and those with odd $t$. The latter ones are in the following denoted 
 as $\tilde{\mathbb{Y}}_{lmt}^c$. According to Eqs.~(\ref{eq:expcom}) 
 and (\ref{eq:exprel}) only even powers of the Cartesian c.m.\ and rel.\ 
 coordinates occur in the Taylor expansions of $\oper{v}_{\rm c.m.}$ 
 and $\oper{v}_{\rm rel.}$ and thus only $\mathbb{Y}_{lmt}^c$ has 
 to be evaluated in those cases. However, the coupling term 
 $\oper{W}$ contains additionally odd powers of 
 $R_c$ and $r_c$ and thus requires also the calculation of 
 $\tilde{\mathbb{Y}}_{lmt}^c$.

 First, the calculation of the even-order coefficients 
 $\mathbb{Y}_{lmt}^x$, $\mathbb{Y}_{lmt}^y$, and 
 $\mathbb{Y}_{lmt}^z$ will be considered. They contain the 
 functions $F_t^c(\theta,\,\phi)$ that are equal to
 $\cos^{2t}(\phi)\sin^{2t}(\theta)$ ($c=x$),
 $\sin^{2t}(\phi)\sin^{2t}(\theta)$ ($c=y$), and 
 $\cos^{2t}(\theta)$ ($c=z$). Consider the integral for
 $\ds\mathbb{Y}_{lmt}^x$. The derivation of $\ds\mathbb{Y}_{lmt}^x$
 is simplified by applying the Euler formula for 
 $\ds\cos^{2t}(\phi)$ and making use of Eq.~(1.111)
 in Ref.~\cite{gen:grad07}. The introduction of the new integration 
 variable $\xi=\cos{(\theta)}$ transforms the integration
 limits in~(\ref{eq:sphercoef}) from $[0,\pi]$ of $\theta$ to 
 $[-1,1]$ for $\xi$. Clearly, the integral is non-zero only, if the 
 integrand is symmetric in the $[-1,1]$ interval. At this point it 
 is important to note that the associated Legendre function  
 $\ds P_l^m$ is even, if the $l+|m|$ sum is even, and odd 
 otherwise. Since the summation index $k$ in Eq.~(1.111) from 
 Ref.~\cite{gen:grad07} is an integer that lies in the interval 
 $\ds 0\le k\le2t$, the relation $\ds -2t\le m\le 2t$ is valid. Hence,
 the $m$ 
 index is always even. Therefore, the integral is non-zero only for 
 even values of $l$. Additionally, there are, of course, the natural 
 restrictions on the $l$ and $m$ coefficients, i.\,e., $\ds l\ge 0$ 
 and $\ds |m|\le l$. Another important fact is that the functions  
 $\ds P_{l>2t}^{|m|\le l,|m| \le 2t}(\xi)$ are oscillatory in the
 interval $[-1,1]$ and the symmetry of the integrand causes the
 contribution of negative and positive parts to cancel out, leading to
 a vanishing integral. Hence, one more restriction on $l$ is 
 $\ds l\le 2t$. Additionally, Eq.~(7.132.1) for the integral over the
 associated Legendre function together with Eqs.~(8.339.2),~(8.339.3)
 and~(8.331.1) (all from Ref.~{\cite{gen:grad07}})  were used 
 in the calculations. 

 Summarizing all the steps mentioned above and
 collecting the indices that do not give trivial zero contributions, the
 analytical form of $\ds \mathbb{Y}_{lmt}^x$ can be given as  
 \begin{multline}
       \ds \mathbb{Y}_{lmt}^x = 
                (-1)^{\frac{l+m}2}
                2^{-\frac{m}2-t+2}
                A_{l-m}\, \pi\,
                {2t\choose t+\frac{m}2}
                 \\
        \ds    \times\, \frac{(t-\frac{m}2)!(t+\frac{m}2)!
               (l-m-1)!!}{(t-\frac{l}2)!
               (\frac{l}2+\frac{m}2)!
               (2t+l+1)!!}\, , 
                \\
               l,m \text{ even,}\; 
               -2t\le m\le 2t,\,\,|m|\le l,\,\, l\le 2t \, .
     \label{eq:yx}
 \end{multline}
 The derivation of the $\mathbb{Y}_{lmt}^y$ coefficient is
 similar to the one for $\mathbb{Y}_{lmt}^x$ and results in 
  \begin{equation}
     \ds \mathbb{Y}_{lmt}^y=
        (-1)^{\frac{m}2}
        \mathbb{Y}_{lmt}^x
        \label{eq:yy}
  \end{equation}
 with the same constraints for the indices as for
 $\mathbb{Y}_{lmt}^x$. In order to derive the
 $\ds \mathbb{Y}_{lmt}^z$ coefficients, Eqs.~(7.231.1) 
 and~(8.752.2) from Ref.~\cite{gen:grad07}  were additionally
 used. This gives 
 \begin{gather}
   \ds  \mathbb{Y}_{lmt}^z = A_{l0}\,
            (-1)^{\frac{l}2}\pi\delta_{m,0}
            \frac{2^{\frac{l}2+2}(2t-1)!!}
            {(l+2t+1)!!}\prod\limits_{i=0}^{l/2-1}(-t+i)
            \, ,
       \nonumber \\
        l \text{ even,}\,\, l\le 2t \, ,
       \label{eq:yz}
 \end{gather}
 where $\delta_{i,j}$ is Kronecker's delta.

 Finally, the odd-order coefficients 
 $\tilde{\mathbb{Y}}_{lmj}^x$,
 $\tilde{\mathbb{Y}}_{lmj}^y$, and
 $\tilde{\mathbb{Y}}_{lmj}^z$ have to be calculated 
 efficiently. They contain the $F_t^c(\theta\, ,\phi)$ 
 functions that are equal to 
 $\cos^{2j+1}(\phi)\sin^{2j+1}(\theta)$,
 $\sin^{2j+1}(\phi)\sin^{2j+1}(\theta)$, and
 $\cos^{2j+1}(\theta)$, respectively. 
 The $\ds \tilde{\mathbb{Y}}_{lmt}^{c}$ coefficients can also be
 calculated analytically. Consider, for example, the integral for
 $\tilde{\mathbb{Y}}_{lmj}^x$. Application of the Euler formula 
 for the $\ds \cos^{2j+1}(\phi)$ term and use of Eqs.~(1.111)
 and~(7.231.2) from Ref.~\cite{gen:grad07}  leads to
 \begin{multline}  
  \ds  \tilde{\mathbb{Y}}_{lmj}^x =
      (-1)^{\frac{l+m}2}
       \ds 2^{ \frac{3-2j-m}{2}}
       A_{l-m}
       {2j+1 \choose j+\ds \frac{m+1}{2} }
       {}
        \\
       {}
       \times\,
       \pi\,
       \frac{
       \left(\frac{2j+1-m}2\right)!
       \left(\frac{2j+1+m}2\right)!
       (l-m-1)!!
       }
       {\left(\frac{2j+1-l}2\right)!
        \left(\frac{m+l}2\right)!
       (2j+l+2)!!} \, ,
       {}
       \\
       {}
       l, m \text{ are odd,}\,\, -2j-1\le m\le 2j+1,
       \,\, l \le 2j+1 \, .
      \label{eq:ypx}
  \end{multline}
 The derivation of $\tilde{\mathbb{Y}}_{lmj}^y$ is also similar to the 
 one of  $\tilde{\mathbb{Y}}_{lmj}^x$ and results in
  \begin{eqnarray}
  \ds 
      \tilde{\mathbb{Y}}_{lmj}^y =
      {\rm i}\, \ds (-1)^{\frac{m-3}2}
      \tilde{\mathbb{Y}}_{lmj}^x\, 
      \label{eq:ypy}
  \end{eqnarray}
 with the same constraints on the indices as for
 $\tilde{\mathbb{Y}}_{lmj}^x$. Finally, $\tilde{\mathbb{Y}}_{lmj}^z$
 is given by
  \begin{equation}
    \ds   \tilde{\mathbb{Y}}_{lmj}^z 
           = A_{l0}
           4\pi\delta_{m,0} (-2)^{\frac{l-1}2}
          \frac{ (2j+1)!!}
          {(2j+l+2)!!} \prod\limits_{i=0}^
          {\frac{l-3}2} (-j+i)
        \label{eq:ypz}
  \end{equation}

 \subsubsection{Final form of the Hamiltonian}
 \label{subsubsec:finalH}

 The final expression for the optical-lattice potential is 
 obtained by inserting $F_t^c(\theta,\phi)$ as defined in
 Eq.~(\ref{eq:sphharexp}) for the angular part of the polynomials of 
 $\vec{R}$ and $\vec{r}$ occurring in 
 Eqs.~(\ref{eq:expcom}), (\ref{eq:exprel}), and~(\ref{eq:expcoupl}). 
 For the three terms
 \begin{multline}
      \ds \oper{v}_{\rm c.m.}
          (R,\Theta,\Phi) =
          -\frac12 \sum\limits_{s=1}^2
          \sum\limits_{c=x,y,z}
          V^s_c \sum\limits_{k=1}^{n}
          \mathbb{C}_{0kcs}^{\text{cos}}
          \, R^{2k}
          {}
          \\
          {}
          \times\, 
          \sum\limits_{L=0,\{2\}}^{2k}
          \sum\limits_{M=-L,\{2\}}^{L}
          \mathbb{Y}_{LMk}^c
          Y_L^M(\Theta,\Phi) \, ,
          \label{eq:oltayexpcom}
 \end{multline}
 \begin{multline}
    \label{eq:oltayexprel}
      \ds
          \oper{v}_{\rm rel.}
          (r,\theta,\phi)=
          -\frac12 \sum\limits_{s=1}^2
          \sum\limits_{c=x,y,z}
          V^s_c \sum\limits_{t=1}^{n}
          \mathbb{C}_{t0cs}^{\text{cos}}
          \, r^{2t}
          {}
          \\
          {}
          \times\, 
          \sum\limits_{l=0,\{2\}}^{2t}
          \sum\limits_{m=-l,\{2\}}^{l}
          \mathbb{Y}_{lmt}^c
          Y_l^m(\theta,\phi) \, ,
 \end{multline}
and
 \begin{multline}
     \ds \oper{W}
         (R,\Theta,\Phi,r,\theta,\phi) =
         \frac12 \sum\limits_{s=1}^2
         \sum\limits_{c=x,y,z}
         V^s_c
         {}
          \\
         {} 
         \times \,
         \left((-1)^{\eta_s}
         \sum\limits_{j=0}^{n-1}
         \sum\limits_{i=0}^{n-1-j}
         \mathbb{C}_{ijcs}^{\text{sin}}
         R^{2i+1} r^{2j+1}\right. 
         {}
          \\
         {}
         \times\, \sum\limits_{l=1,\{2\}}^{2j+1}
         \left[\tilde{\mathbb{Y}}_{l0j}^c
         Y_l^0(\theta,\phi)+
         \sum\limits_{m=-l,\{2\}}^{l}
         \tilde{\mathbb{Y}}_{lmj}^c
         Y_l^m(\theta,\phi) \right]
         {}
          \\
         {}
         \times\, \sum\limits_{L=1,\{2\}}^{2i+1}
         \left[\tilde{\mathbb{Y}}_{L0i}^c
         Y_L^0(\Theta,\Phi)+
         \sum\limits_{M=-L,\{2\}}^{L}
         \tilde{\mathbb{Y}}_{LMi}^c
         Y_L^M(\Theta,\Phi) \right]
         {}
          \\
         {}
         -\, \sum\limits_{t=1}^{n}
         \sum\limits_{k=1}^{n-t}
         \mathbb{C}_{tkcs}^{\text{cos}}
         R^{2k} r^{2t} 
         \sum\limits_{l=0,\{2\}}^{2t}
         \sum\limits_{m=-l,\{2\}}^{l}
         \mathbb{Y}_{lmt}^c
         Y_l^m(\theta,\phi)
         {}
          \\
         {}
          \times\, 
          \left. \sum\limits_{L=0,\{2\}}^{2k}
         \sum\limits_{M=-L,\{2\}}^{L}
         \mathbb{Y}_{LMk}^c
         Y_L^M(\Theta,\Phi)
           \right)
         \label{eq:wsphharm}
  \end{multline}
 is found where, e.\,g., $\sum\limits_{l=0,\{2\}}^{2t}$ stands for 
 $\sum\limits_{l=0,2,4,\dots}^{2t}$. In Eq.~(\ref{eq:wsphharm}) 
 $\tilde{\mathbb{Y}}_{l0j}^x=\tilde{\mathbb{Y}}_{l0j}^y=0$ 
 and $\tilde{\mathbb{Y}}_{lmj}^z=0$ for $m\ne 0$ is implied.

 As a result of adopting spherical c.m.\ and rel.\ coordinates 
 the Hamiltonian
 \begin{align}
  \ds \oper{H}(r,\theta,\phi, R,\Theta, \Phi) 
      &= \oper{h}_{\rm rel.}(r,\theta,\phi)
       + \oper{h}_{\rm c.m.}(R,\Theta,\Phi)
  \nonumber \\
      & \quad +\oper{W}(r,\theta,\phi, R,\Theta, \Phi)
      \label{eq:hamcmrelspfr}
 \end{align}
 is obtained with 
 \begin{align}
   \ds  \oper{h}_{\rm c.m.}(R,\Theta,\Phi) =
       - \frac{1}{2{\rm M}} 
                \left[
        \frac{\partial^2}{\partial R^2} \right. 
       &       \left.
       +  \frac{2}{R} \frac{\partial}{\partial R} 
        - \frac{{{\bf \hat{I}}^2_{\rm c.m.} }}{R^2} 
               \right] 
  \nonumber \\
        &\hspace{0.2cm}
       +\oper{v}_{\rm c.m.}(R,\Theta,\Phi)
 \label{eq:orbitcom}
 \end{align}
and
 \begin{align} 
   \ds \oper{h}_{\rm rel.}(r,\theta,\phi) = 
       -\frac{1}{2\mu}
                \left[
        \frac{\partial^2}{\partial r^2} \right.  
       &       \left.
       +  \frac{2}{r} \frac{\partial}{\partial r} 
        - \frac{{{\bf \hat{I}}^2_{\rm rel.} }}{r^2} 
               \right] 
        +  \oper{u} (r) 
  \nonumber \\
        &\hspace{0.2cm}+
        \oper{v}_{\rm rel.}(r,\theta,\phi) \, .
        \label{eq:hrelsph}
  \end{align}
 In these equations $\oper{\bf I}_{\rm c.m.}$ and 
 $\oper{\bf I}_{\rm rel.}$ are the operators of angular momentum, 
 $\mu=m_1m_2/(m_1+m_2)$ is the reduced mass, and ${\rm M}=m_1+m_2$ is
 the total mass of the two particles. 

 The key achievement is that now all terms in the Hamiltonian are at most a sum 
 over products of functions that depend only on a single coordinate, 
 i.\,e., $f_1(R)f_2(\Theta)f_3(\Phi)g_1(r)g_2(\theta)g_3(\phi)$. 
 As a result all required integrals are at most products of 
 one-dimensional integrals.

 \subsubsection{Finite and $\cos^2$ lattices}
 \label{subsubsec:finite}

\begin{figure}[!ht]
 \centering
\includegraphics[width=0.4\textwidth]{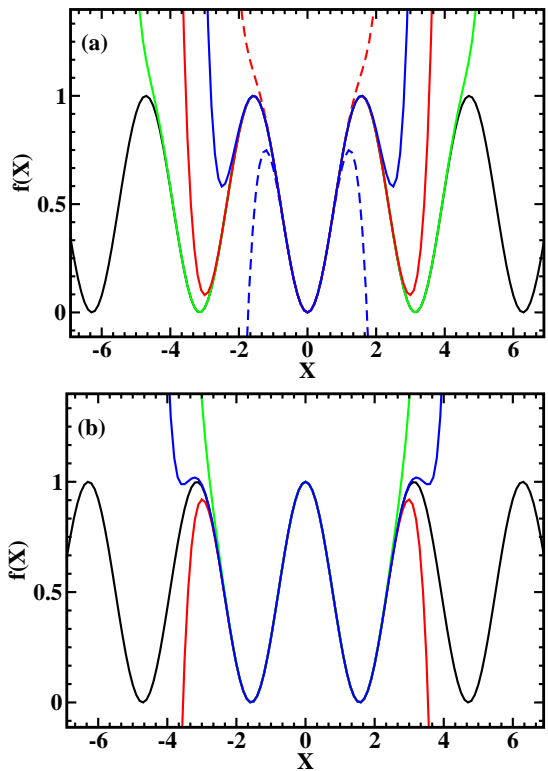}
 \caption[Double-well and triple-well traps]
     {
     (a) The $\sin^2(x)$ function (black solid) together with
         the 2nd- (blue dashes), 3rd- (red dashes),
	 5th- (blue solid), 7th- (red solid) and
	 11th-order (green solid)  expansion of the Taylor series.
     (b) The $\cos^2(x)$ function (black) together with 
         the 6th- (green), 7th- (red) and 8th-order (blue) expansion 
         of the Taylor series.
     }
     \label{fig:sincosexp}
 \end{figure}
 For practical reasons, the infinite Taylor expansion of the 
 $\sin^2$ potential has to be truncated in a calculation. In 
 such a situation convergence of the results with respect 
 to the expansion length is usually aimed for. However, in the 
 present case some caution has to be applied in such a study and the 
 interpretation of its outcome \cite{cold:gris09}. 
 The problem is due to the fact that 
 the Taylor expansion is performed around the origin. Hence the 
 $\sin^2$ lattice is best described close to the origin, while 
 the deviations increase with increasing distance from it. 
 This is illustrated in Fig.~\ref{fig:sincosexp}(a) in which 
 the $\sin^2$ lattice (along one coordinate) 
 is compared to Taylor expansions of  
 different order. While the 2nd-order expansion works already 
 very well at and close to the origin, even the barrier height 
 of the central well is not correctly described. On the other hand, 
 the 3rd-order expansion that includes polynomials up to 6th order 
 and may thus be called sextic potential agrees very well with the 
 central well of the $\sin^2$ potential. The sextic potential is 
 thus a very good choice for the investigation of the effects of 
 anharmonicity on (tightly) bound states in a single site of an optical
 lattice \cite{cold:gris09}. While increasing the order of the 
 expansion by considering, e.\,g., the 5th- or 7th-order expansion 
 improves the agreement with the $\sin^2$ potential further away from 
 the origin, a problem occurs. The resulting potential possesses 
 three wells, but the outer ones have a depth that differs
 from the correct one, as is also illustrated in
 Fig.~\ref{fig:sincosexp}(a). As a consequence, non-physical
 resonances may occur due to tunneling. Hence, a simple convergence 
 study in which the Taylor expansion is expanded order by order 
 is problematic. In fact, an even more severe problem is observed 
 for all even-order expansions like the already discussed 2nd-order 
 one. Those expansions tend to $-\infty$ for $x$ going to 
 either $-\infty$ or $+\infty$ (or even in both cases as for the 
 shown 2nd-order expansion).  As a consequence, an infinite number 
 of unphysical negative-energy states occur.
    
 In conclusion, the present approach that is based on a Taylor 
 expansion of the optical-lattice potential is especially suitable 
 for describing finite optical lattices. With a judicious choice of 
 the expansion of the $\sin^2$ potential the physics of single- 
 triple-, or higher multiple-well potentials can be well described. 
 For example, the $11$th-order expansion also shown in 
 Fig.~\ref{fig:sincosexp}(a) provides a very good description 
 of the physics in a triple-well potential \cite{cold:schn09}. 
 Clearly, even an expansion like the 5th-order one may be useful, 
 if an asymmetric potential with different depths of the wells 
 should be considered.
 Furthermore, with a sufficiently large number of wells even the 
 physics of a complete optical lattice can be described in which the 
 continuum states (or transitions to them) are involved. In that 
 case convergence with respect to the number of wells (and not 
 with respect to the expansion length) has to be achieved, since the 
 true continuum is replaced by a correspondingly discretized 
 spectrum. In fact, it may also be reminded that in most experiments 
 involving ultracold atoms in optical lattices there is an additional 
 confining potential and thus the whole (relevant) spectrum can be 
 discrete.  

 An evident limitation of the $\sin^2$ lattice and its Taylor expansion 
 discussed so far is that only finite lattices with an odd number of 
 wells can easily be described. Clearly, in many situations also 
 finite lattices with an even number of potential wells are 
 of interest. The most prominent example is certainly the double-well 
 potential that is also of special interest for quantum-information 
 studies. The physics of few atoms in (one-dimensional) double-well 
 potentials was recently studied, e.\,g., in \cite{cold:zoel08}. 
 The most straightforward extension of the present approach towards 
 such potentials is by considering a Taylor expansion of the 
 $\cos^2$ (or $\pi/2$-shifted $\sin^2$) potential 
  \begin{equation}
    \ds
     \oper{{V}}^{\cos}(\vec{R},\vec{r}) =
     \sum\limits_{s=1}^2 \sum\limits_{c=x,y,z} 
     V_c^s \sin^2(k_c c_s + \frac{\pi}2) \, .
  \end{equation}
 Using trigonometric relations the optical-lattice potential can be
 written in the more suitable form
  \begin{multline}
    \ds
         \oper{V}^{\cos}(\vec{R},\vec{r}) =  
         \frac12 \sum\limits_{s=1}^2  
         \sum\limits_{c=x,y,z}
         V^s_c [1 + 
         (-1)^{s}
         \sin (2k_c R_c) 
         {}
          \\
         {}
         \times\,
         \sin (2k_c r_c \mu_{\eta_s})
         +\,
         \cos (2k_c R_c) 
         \cos (2k_c r_c \mu_{\eta_s})]\, .
         \label{eq:olcomrmgen}
   \end{multline}
 After derivations similar to the case of the $\sin^2$ potential the
 splitting of the optical lattice into the c.m.\ and rel.\ motion in the
 Cartesian frame yields
%
%   \begin{equation}
%     \ds  \oper{v}_0^{\cos} =
%          \sum\limits_{s=1}^2
%          \sum\limits_{c=x,y,z} V^s_c \, ,
%          \label{eq:cos2a}
%   \end{equation}
% %
%   \begin{equation}
%         \oper{v}_{\rm c.m.}^{\cos}
%         (\vec{R}) =
%         \frac12 \sum\limits_{s=1}^2
%         \sum\limits_{c=x,y,z}
%         V^s_c \sum\limits_{k=1}^{n}
%         \mathbb{C}_{0kcs}^{\text{cos}}
%         R_c^{2k} \, ,
%         \label{eq:cos2b}
%   \end{equation}
% %
%   \begin{equation}
%         \oper{v}_{\rm rel.}^{\cos}
%         (\vec{r}) =
%         \frac12 \sum\limits_{s=1}^2
%         \sum\limits_{c=x,y,z}
%         V^s_c \sum\limits_{t=1}^{n}
%         \mathbb{C}_{t0cs}^{\text{cos}}
%         r_c^{2t} \, ,
%         \label{eq:cos2c}
%   \end{equation}
% %
%   \begin{multline}
%          \oper{W}^{\cos}
%              (\vec{R},\vec{r}) =
%          \frac12 \sum\limits_{s=1}^2
%              \sum\limits_{c=x,y,z}
%              V^s_c 
%              \Biggl[(-1)^{s}
%              \sum\limits_{j=0}^{n-1}
%               {}
%               \\
%               {}
%              \times\,
%              \sum\limits_{i=0}^{n-1-j}
%              \mathbb{C}_{ijcs}^{\text{sin}}
%              R_c^{2i+1} r_c^{2j+1} +
%              \sum\limits_{t=1}^{n}
%              \sum\limits_{k=1}^{n-t}
%              \mathbb{C}_{tkcs}^{\text{cos}}
%              R_c^{2k} r_c^{2t} \Biggr] 
%              \label{eq:cos2d}
%   \end{multline}
%
  \begin{align}
    \ds  \oper{v}_0^{\cos} &=
         \sum\limits_{s=1}^2
         \sum\limits_{c=x,y,z} V^s_c \label{eq:cos2a}\\
        \oper{v}_{\rm c.m.}^{\cos}
        (\vec{R}) &=
        \frac12 \sum\limits_{s=1}^2
        \sum\limits_{c=x,y,z}
        V^s_c \sum\limits_{k=1}^{n}
        \mathbb{C}_{0kcs}^{\text{cos}}
        R_c^{2k}       \label{eq:cos2b}\\
        \oper{v}_{\rm rel.}^{\cos}
        (\vec{r}) &=
        \frac12 \sum\limits_{s=1}^2
        \sum\limits_{c=x,y,z}
        V^s_c \sum\limits_{t=1}^{n}
        \mathbb{C}_{t0cs}^{\text{cos}}
        r_c^{2t}         \label{eq:cos2c}\\
         \oper{W}^{\cos}
             (\vec{R},\vec{r}) &=
         \frac12 \sum\limits_{s=1}^2
             \sum\limits_{c=x,y,z}
             V^s_c 
             \Biggl[(-1)^{s}
             \sum\limits_{j=0}^{n-1}
                \label{eq:cos2d}  \\
             &\hspace{-1.2cm} \times\,
             \sum\limits_{i=0}^{n-1-j}
             \mathbb{C}_{ijcs}^{\text{sin}}
             R_c^{2i+1} r_c^{2j+1} +
             \sum\limits_{t=1}^{n}
             \sum\limits_{k=1}^{n-t}
             \mathbb{C}_{tkcs}^{\text{cos}}
             R_c^{2k} r_c^{2t} \Biggr]   \nonumber         
  \end{align}
 where the constant term $\oper{v}_0^{\cos}$ appears that 
 was zero for the $\sin^2$ case.
 Equations~(\ref{eq:cos2a})-(\ref{eq:cos2d}) are almost analogous to
 Eqs.~(\ref{eq:expcom})-(\ref{eq:expcoupl}) for the $\sin^2$-like
 potential. In Fig.~\ref{fig:sincosexp}(b) the $\cos^2$ lattice 
 (along one coordinate) is shown together with Taylor expansions 
 of different order. In this case, an even-order expansion
 should 
 be used to avoid unphysical negative energy states. For example, 
 while the 6th-order expansion provides a rather good representation 
 of two neighbor wells in an optical lattice, the 7th-order expansion 
 leads to negative energy states and does in fact even not represent 
 the outer potential barriers properly. The 8th-order expansion 
 is again rather good, but leads already to small artificial side 
 minima. 
 
 Since the adopted expansions are independent for the three orthogonal 
 directions $x$, $y$, and $z$, the present approach is capable of 
 describing two particles in any combination of different lattices 
 along those three directions. While, e.\,g., in \cite{cold:schn09} 
 a true triple well was described using the 11th-order expansion of $\sin^2$ 
 in one and a harmonic (1st-order) expansion in the two other 
 directions, alternatively an array of $3\times 3 \times 3$ 
 lattice sites could be described equally well.

 \section{Exact diagonalization}
 \label{sec:hamandwf}

 \subsection{Schr\"odinger equations}
 \label{subsec:schroedinger}

 After having formulated the optical lattice potential in a
 suitable form, the solution of the eigenvalue problem is described in
 the following. The Schr\"odinger equation with the Hamiltonian of (\ref{eq:hamcmrelspfr})
 \begin{equation}
 \label{eq:fullSchroed}
 \oper{H}\,|\Psi_i\rangle = \eci_i \,|\Psi_i\rangle\, ,
 \end{equation}
 is solved by expanding $\Psi$ in terms of configurations, 
 \begin{equation}
   \Psi_i(\vec{R},\vec{r}\,) = \sum_{\kappa} \, C_{i,\kappa} \, 
                                  \Phi_\kappa (\vec{R},\vec{r}\,) \, .
 \label{eq:Psi}
 \end{equation}
 The {\it configurations} 
 \begin{equation}
      \Phi_\kappa(\vec{R},\vec{r}) =
              \varphi_{i_\kappa}(\vec{r}\,) \, \psi_{j_\kappa}(\vec{R}\,)
 \label{eq:config}
 \end{equation}
 are products of the eigenfunctions of the Hamiltonians 
 of rel.\ and c.m.\ motions, respectively, i.\,e., $\varphi$ 
 and $\psi$ are solutions of
 \begin{equation}
   \oper{h}_{\rm rel.} \,\ket{\varphi_{i}} = \epsilon_{i}^{\rm rel.} \,
                                           \ket{\varphi_{i}} \, 
   \label{eq:schrel}
 \end{equation}
 and
 \begin{equation}
   \oper{h}_{\rm c.m.} \,\ket{\psi_{j}} =\epsilon_{j}^{\rm c.m.} \, 
                                           \ket{\psi_{j}}.
   \label{eq:schcom}
 \end{equation}
 Finally, the wavefunctions of rel.\ and c.m.\ motion that we 
 denote as {\it orbitals} (in formal analogy to the one-particle 
 solutions in electronic-structure calculations) are expressed 
 in basis functions $\tilde{\varphi}$ and $\tilde{\psi}$ that 
 are products of $B$ splines and spherical harmonics $Y_l^m$ 
 for describing the radial and the angular parts, respectively,
 \begin{align}
       \varphi_{i} (\vec{r}) &= 
           \sum\limits_{\mathbf{a}} \tilde{c}^{\rm rel.}_{i,\mathbf{a}} \,
                   \tilde{\varphi}_{\mathbf{a}} 
        \label{eq:wfrel1}
        \\
       & =
        \sum\limits_{\alpha=1}^{N_r}
        \sum\limits_{l=0}^{N_l}
        \sum\limits_{m=-l}^{l}
        \tilde{c}^{\rm rel.}_{i, \alpha l m} \,r^{-1}\,B_{\alpha} (r)
         Y_l^m (\theta,\phi)
        \label{eq:wfrel2}
 \end{align}
and
 \begin{align}
      \psi_{j} (\vec{R}) &=
           \sum\limits_{\mathbf{b}} \tilde{c}^{\rm c.m.}_{j,\mathbf{b}} \,
                        \tilde{\psi}_{\mathbf{b}} 
        \label{eq:wfcom1}
        \\
      & =
       \sum\limits_{\beta=1}^{N_R}
       \sum\limits_{L=0}^{N_L}
       \sum\limits_{M=-L}^{L}
       \tilde{c}^{\rm c.m.}_{j, \beta L M} \,R^{-1}\, B_{\beta}(R)
        Y_L^M (\Theta,\Phi)  \, .
       \label{eq:wfcom2}
 \end{align}
 In Eqs.~(\ref{eq:wfrel1}) and (\ref{eq:wfcom1}), we introduced the 
 compact indices $\mathbf{a} \equiv \alpha, l, m$ and 
 $\mathbf{b} \equiv \beta, L, M$. 

 A specific basis set is 
 characterized by the upper limits of angular momentum $N_l$ and $N_L$ 
 in the spherical-harmonic expansions and the numbers $N_r$ and $N_R$ 
 of $B$ splines used in the expansions in Eqs.~(\ref{eq:wfrel2}) and 
 (\ref{eq:wfcom2}) as well as their order $k_{\rm rel.}$ (and
 $k_{\rm c.m.}$) and knot 
 sequences \cite{bsp:boor78,bsp:bach01}. The knot sequences define  
 the ranges of $r$ and $R$ in which the wave functions are calculated, 
 the so-called {\it box}, though it is, in fact, often a sphere as 
 in the present case. If the box is chosen sufficiently large for a 
 given finite trapping potential, all wavefunctions will have decayed 
 before reaching the box boundaries. Otherwise, an artificial 
 discretization occurs, if a zero-boundary condition at the wall 
 of the box is enforced by removing the last $B$ spline. In this case 
 an investigation of the convergence of the results with respect to 
 the box size has to be performed. 

 The insertion of the expansions for $\varphi$ in  
 (\ref{eq:wfrel2}) and $\psi$ in (\ref{eq:wfcom2}) 
 into the Schr\"odinger equations (\ref{eq:schrel}) and  
 (\ref{eq:schcom}), respectively, followed by a multiplication with 
 either $\tilde{\varphi}_i^*$ or $\tilde{\psi}_j^*$ 
 (from left) and integration over $\vec{r}$ or $\vec{R}$ leads to 
 generalized matrix eigenvalue problems of the type
 \begin{equation}
   \ds
      \mathbf{h}\mathbf{\tilde{c}}_i=\epsilon_i\, \mathbf{s}
                \mathbf{\tilde{c}}_i\, .
   \label{eq:geneiv}
 \end{equation}
 Their solutions provide the energies $\epsilon_i^{\rm rel.}$ (and
 $\epsilon_j^{\rm c.m.}$) 
 as well as the coefficients $\tilde{c}^{\rm rel.}_{i,\mathbf{a}}$ 
 (and $\tilde{c}^{\rm c.m.}_{j,\mathbf{b}}$). The latter define the rel.\ 
 and c.m.\ orbitals $\varphi$ and $\psi$ according to Eqs.~(\ref{eq:wfrel1}) 
 and (\ref{eq:wfcom1}), respectively. Generalized eigenvalue equations 
 occur due to the non-orthogonality of the $B$ splines. Furthermore, 
 the explicit consideration of the factors $r^{-1}$ and $R^{-1}$ 
 transforms the radial part of the Schr\"odinger equations into 
 effective one-dimensional ones by removing the $\partial/\partial R$ 
 and $\partial/\partial r$ terms in Eqs.~(\ref{eq:orbitcom}) and 
 (\ref{eq:hrelsph}). As a consequence, the diagonalizations provide in fact 
 the solutions $r\varphi$ and $R\psi$ from which $\varphi$ and $\psi$ 
 can, of course, easily be obtained. Since $r\varphi$ and $R\psi$ 
 vanish for $r\rightarrow 0$ and $R\rightarrow 0$, respectively, 
 this additional boundary condition is implemented by removing the 
 first $B$ spline. Together with the corresponding boundary condition 
 at the outer box boundaries, the summations in Eqs.~(\ref{eq:wfrel2}) 
 and (\ref{eq:wfcom2}) change into $\sum_{\alpha=2}^{N_r-1}$ 
 and $\sum_{\beta=2}^{N_r-1}$, respectively. In fact, the actual 
 implementation of the code is flexible with respect to different 
 choices of the boundary conditions at the origin of rel.\ and c.m.\ 
 motions, but in the following only the standard use based on the reduced 
 summation limits is considered explicitly for reasons of better 
 readability.
 
 Once the eigenvectors $\varphi$ and $\psi$ are obtained, a set 
 of configurations $\Phi$ is built according to Eq.~(\ref{eq:config}). 
 Again, insertion of the expansion for $\Psi$ in Eq.~(\ref{eq:Psi}) 
 into the Schr\"odinger equation (\ref{eq:fullSchroed}), multiplication 
 with $\Phi_k^*$ (from left), and integration over $\vec{r}$ and 
 $\vec{R}$ yields the matrix eigenvalue equation
 \begin{equation}
   \ds
    \mathbf{H}\mathbf{C}_i=\eci_i\, \mathbf{C}_i\, .
   \label{eq:eigenvalue}
 \end{equation}
 Due to the orthonormality of the orbitals $\varphi$ and $\psi$ 
 also the configurations $\Phi$ are orthonormal. Therefore, 
 the overlap matrix is equal to the identity and 
 Eq.~(\ref{eq:eigenvalue}) is an ordinary eigenvalue problem.

 \subsection{Matrix elements}
 \label{subsec:matel}

 In order to set up the matrix eigenvalue problems in 
 Eqs.~(\ref{eq:geneiv}) and (\ref{eq:eigenvalue}), the 
 corresponding matrices
 \begin{eqnarray}
       h_{\mathbf{a},\mathbf{a}'}^{\rm rel.} = 
           \opij{\tilde{\varphi}_\mathbf{a}}{\oper{h}_{\rm rel.}}
                           {\tilde{\varphi}_{\mathbf{a}'}}\, ,
       \label{eq:hrel}
      &\quad&
       s_{\mathbf{a},\mathbf{a}'}^{\rm rel.} = 
           \braket{\tilde{\varphi}_\mathbf{a}}
                   {\tilde{\varphi}_{\mathbf{a}'}} \, ,
       \label{eq:srel} \\
       h_{\mathbf{b},\mathbf{b}'}^{\rm c.m.} = 
           \opij{\tilde{\psi}_\mathbf{b}}{\oper{h}_{\rm c.m.}}
                           {\tilde{\psi}_{\mathbf{b}'}}\, ,
       \label{eq:hcom}
      &\quad&
       s_{\mathbf{b},\mathbf{b}'}^{\rm c.m.} = 
           \braket{\tilde{\psi}_\mathbf{b}}
                   {\tilde{\psi}_{\mathbf{b}'}}  
       \label{eq:scom}
  \end{eqnarray}
 and
 \begin{equation}
     H_{\kappa,\kappa'} = 
         \opij{\Phi_\kappa}{\oper{H}}{\Phi_{\kappa'}}
     \label{eq:Hkk}
 \end{equation}
 have to be set up. As already mentioned, the overlap matrix elements 
 between configurations are trivial,
 \begin{multline}
    S_{\kappa,\kappa'} = \braket{\Phi_\kappa}{\Phi_{\kappa'}}
        {}
	\\
	{}
        = \braket{\varphi_{i_\kappa}\psi_{j_\kappa}}
                     {\varphi_{i_{\kappa'}}\psi_{j_{\kappa'}}}
        = \delta_{i_{\kappa{\vphantom{'}}},i_{\kappa'}} 
	  \, \delta_{j_{\kappa\vphantom'},j_{\kappa'}}
        = \delta_{\kappa,\kappa'} \, . 
 \end{multline}

 For convenience, the integrals over $B$ splines and their 
 derivatives are denoted as
 \begin{equation}
   \ds \mathbb{B}_{\partial^\mu\alpha\,\partial^\nu\alpha'}^{\lambda} = 
   \int\limits_0^{\infty}\,dr\, r^{\lambda}
            \frac{\partial^\mu B_{\alpha}(r)}{\partial r^\mu}
            \frac{\partial^\nu B_{\alpha'}(r)}{\partial r^\nu}\, .
  \label{eq:bsplinterrel}
 \end{equation}
 Furthermore, the index $\lambda$ and the orders of the derivatives 
 $\mu$ (or $\nu$) 
 are omitted for $\mu=0$ ($\nu=0$). For example, one has 
 $\mathbb{B}_{\alpha\alpha'}\equiv
 \mathbb{B}_{\partial^0\alpha\partial^0\alpha'}^{0}$. 
 Additionally, it is reminded that the character $\alpha$ is reserved 
 for rel.\ motion matrix elements and $\beta$ for c.m.\ elements. Hence, the 
 corresponding notation for the c.m.\ integral over $B$ splines 
 analogous to Eq.~(\ref{eq:bsplinterrel}) is  
 \begin{equation}
   \ds \mathbb{B}_{\partial^\mu\beta\,\partial^\nu\beta'}^{\lambda} = 
   \int\limits_0^{\infty}\,dR\, R^{\lambda}
            \frac{\partial^\mu B_{\beta}(R)}{\partial R^\mu}
            \frac{\partial^\nu B_{\beta'}(R)}{\partial R^\nu}\, .
  \label{eq:bsplintercom}
 \end{equation}
 Since $B$ splines are polynomials, the integrals $\mathbb{B}$ 
 can be calculated exactly by means of Gauss-Legendre quadrature.  
 Due to the compactness (finite local support) of the $B$ splines 
 the integration limits are in fact finite. If the two involved $B$ 
 splines do not possess a common interval where both of them 
 are non-zero, the integral vanishes. Therefore, 
 only a very limited number of integrals has to be calculated and 
 the resulting overlap and Hamiltonian matrices are sparse.  
 In the following, all integrals that occur in the calculation 
 are discussed individually.

 \subsubsection{Overlap}

 The overlap matrices between the basis functions $\tilde{\varphi}$ 
 and $\tilde{\psi}$ are not equal to the identity matrix, but
 \begin{equation}
   \ds s_{\mathbf{a},\mathbf{a'}}^{\rm rel.} = \mathbb{B}_{\alpha\alpha'}
	\ds \int_\Omega\, d\Omega\,
	    {Y_{l}^{m}}^*
            (\theta,\phi)Y_{l'}^{m'}
            (\theta,\phi) 
	    = \mathbb{B}_{\alpha\alpha'}
            \delta_{ll'}\delta_{mm'}
            \label{eq:overlaprel}
  \end{equation}
 and, similarly,
 \begin{equation}
  \ds s_{\mathbf{b},\mathbf{b'}}^{\rm c.m.} = \mathbb{B}_{\beta\beta'}
              \delta_{LL'}\delta_{MM'}\, .
              \label{eq:overlapcom}
 \end{equation}

 \subsubsection{Kinetic energy}

 Since the basis functions are a product of a radial $B$ spline 
 and a spherical harmonic, the action of the kinetic-energy 
 operator on them is straightforwardly calculated. Using 
 $\ds {{\bf \hat{I}}^2_{\rm rel.} }
  Y_l^m(\theta,\phi)=l(l+1)Y_l^m(\theta,\phi)$  
 and $\ds {{\bf \hat{I}}^2_{\rm c.m.} }
 Y_L^M(\Theta,\Phi)=L(L+1)Y_L^M(\Theta,\Phi)$, one finds 
  \begin{multline}
  \ds
       t_{\mathbf{a},\mathbf{a'}}^{\rm rel.} =
      -\frac{1}{2\mu}
       \mathbb{B}_{\partial^2\alpha\, \alpha'} \,
       \delta_{ll'}\delta_{mm'}
       + \frac{1}{2\mu}l(l+1)\,
       \mathbb{B}^{-2}_{\alpha\alpha'} \,
       \delta_{ll'}\delta_{mm'} {}
        \\
       =  
       {} \frac{1}{2\mu} \left( 
      \mathbb{B}_{\partial^1\alpha\, \partial^1\alpha'}
       + l(l+1)\,
      \mathbb{B}^{-2}_{\alpha\alpha'}
       \right) \delta_{ll'}\delta_{mm'}
    \label{eq:trel}
 \end{multline}
 for the rel.\ motion and analogously 
 \begin{equation}
  \ds
       t_{\mathbf{b},\mathbf{b'}}^{\rm c.m.} = 
       \frac{1}{2{\rm M}} 
       \left( 
	   \mathbb{B}_{\partial^1\beta\, \partial^1\beta'}
	    + L(L+1)\, 
	   \mathbb{B}^{-2}_{\beta\beta'}
       \right)\,
       \delta_{LL'}\delta_{MM'} 
       \label{eq:tcom}
\end{equation}
 for the c.m.\ motion. Note, the second equality in Eq.~(\ref{eq:trel}) 
 as well as Eq.~(\ref{eq:tcom}) have to be (slightly) modified, if 
 non-zero boundary conditions are applied at the origin and the box 
 boundary.

 \subsubsection{Interparticle interaction}

 The matrix elements of the interparticle interaction potential are
\begin{equation}
    \ds 
       u_{\mathbf{a},\mathbf{a'}} =
        \delta_{ll'}\delta_{mm'} \,
        \int\limits_{0}^{\infty}
        \,dr\, \, u (r)
        B_{\alpha}(r) \, B_{\alpha'}(r)\, .
        \label{eq:urel}
  \end{equation}
 The compactness of the $B$ splines turns the semi-indefinite integral 
 into a definite one that has to be calculated only within a small 
 spatial interval in which $B_{\alpha}$ and $B_{\alpha'}$ 
 (and, of course, $u(r)$) are simultaneously non-zero. In contrast 
 to the case of the $\mathbb{B}$ integrals Gauss quadrature is in this 
 case only exact, if $u(r)$ can be expressed in terms of a finite 
 polynomial expansion. In practice, the quadrature converges quite 
 well even with a relative small number of terms. This is again partly 
 due to the fact that it is sufficient, if a polynomial expansion 
 works well piecewise, i.\,e., only within small spatial intervals.

 \subsubsection{Separable part of the trapping potential}

 Using the property
 $\ds {Y_{l_t}^{m_t}}^*(\theta,\phi) = 
  (-1)^{m_t} {Y_{l_t}^{-m_t}}(\theta,\phi)$ the product of two
 spherical harmonics can be expressed as a sum of products between
 a spherical harmonic and the 3j-Wigner symbols, 
 \begin{multline}
   \ds Y_l^m (\theta,\phi) \,
       Y_{l_t}^{m_t}(\theta,\phi) = {}
       \\
       {} \sum\limits_{l_t,m_t} 
       \mathbb{A}^0_{l_t\, l\, l_t}
       \begin{pmatrix} l_t & l & l_t \\ 
       m_t & m & m_t \end{pmatrix}
       \begin{pmatrix} l_t & l & l_t \\ 
       0 & 0 & 0 \end{pmatrix}
       {Y_{l_t}^{m_t}}^*(\theta,\phi) \, .
 \end{multline}
 Here, the coefficient 
 \begin{equation}
   \ds \mathbb{A}^a_{b\,c\,d}=
   (-1)^a
   \sqrt{\frac{(2b+1)(2c+1)(2d+1)}{4\pi}}\, 
 \end{equation}
 was introduced for compactness. The Gaunt
 coefficient may be obtained as~\cite{cold:rasc03,cold:pinc07} 
 \begin{multline}
   \ds \int_\Omega\, d\Omega\,
        Y_l^m (\theta,\phi)
        Y_{l_t}^{m_t}(\theta,\phi)
        {Y_{l'}^{m'}}^* (\theta,\phi) {}
         \\
      {} = \sum\limits_{l_t,m_t} 
        \mathbb{A}^{m_t}_{l_t\, l\, l_t}
        \begin{pmatrix} l_t & l & l_t \\ 
         m_t & m & m_t \end{pmatrix}
        \begin{pmatrix} l_t & l & l_t \\ 
          0 & 0 & 0 \end{pmatrix} {}
         \\
	{} \times \,
        \underbrace{\ds
	\int_\Omega\, d\Omega\,
	{Y_{l_t}^{-m_t}}
        (\theta,\phi){Y_{l'}^{m'}}^*
        (\theta,\phi)}_
        {\ds \delta_{l_tl'}\,\delta_{-m_tm'}} {}
          \\
	{} 
	= \mathbb{A}^{m'}_{l_t\, l\, l'}
        \begin{pmatrix} l_t & l & l' \\ 
        m_t & m & -m' \end{pmatrix}
        \begin{pmatrix} l_t & l & l' \\ 
         0 & 0 & 0 \end{pmatrix} \, .
        \label{eq:gaunt}
 \end{multline}
 Making use of Eq.~(\ref{eq:gaunt}), the angular parts for the matrix
 elements of the trapping potential can be calculated straightforwardly. 
 For the separable (uncoupled) parts  
 \begin{multline}
   \ds  v_{\mathbf{a},\mathbf{a'}}^{\rm rel.} =
            -\frac12 \sum\limits_{s=1}^2
            \sum\limits_{c=x,y,z}
            V^s_c \sum\limits_{t=1}^{n}
            \mathbb{C}_{t0cs}^{\text{cos}}\,
	    \mathbb{B}^{2t}_{\alpha\, \alpha'} 
            \sum\limits_{{l_t}=0,\{2\}}^{2t}
	    {}
	     \\
            {} \times\,
            \sum\limits_{{m_t}=-{l_t},\{2\}}^{{l_t}}
            \mathbb{Y}_{l_t{m_t}t}^c 
            \mathbb{A}^{m'}_{l_t\, l\, l'} {}
            \begin{pmatrix} l_t & l & l' \\ 
            m_t & m & -m' \end{pmatrix}
            \begin{pmatrix} l_t & l & l' \\ 
            0 & 0 & 0 \end{pmatrix}
            \label{eq:vrel} 
 \end{multline}
 and
   \begin{multline}
     \ds  V_{\mathbf{b},\mathbf{b'}}^{\rm c.m.} =
              -\frac12 \sum\limits_{s=1}^2
              \sum\limits_{c=x,y,z}
              V^s_c \sum\limits_{k=1}^{n}
              \mathbb{C}_{0kcs}^{\text{cos}}
	      \, \mathbb{B}^{2k}_{\beta\, \beta'} 
	      {}
	       \\
              {} \times \,
              \sum\limits_{{L_k}=0,\{2\}}^{2k}
              \sum\limits_{{M_k}=-{L_k},\{2\}}^{{L_k}}
              \mathbb{Y}_{L_k{M_k}k}^c 
	      \mathbb{A}^{M'}_{L_k\, L\, L'}    
	      {}
               \\
              {} \times\,
              \begin{pmatrix} L_k & L & L' \\ 
              M_k & M & -M' \end{pmatrix}
              \begin{pmatrix} L_k & L & L' \\ 
              0 & 0 & 0 \end{pmatrix} \, .
              \label{eq:vcom}   
   \end{multline}   
 are found for the rel.\ and c.m.\ matrix elements, respectively. With the 
 aid of Eqs.~(\ref{eq:overlaprel}), (\ref{eq:trel}), (\ref{eq:urel}), 
 and (\ref{eq:vrel}) the rel.\ overlap and Hamiltonian matrices in 
 Eqs.~(\ref{eq:srel}) are obtained. Insertion into 
 Eq.~(\ref{eq:geneiv}) and subsequent diagonalization yields the uncoupled 
 eigenenergies and eigenfunctions of the rel.\ motion, as discussed 
 above. Analogously, Eqs.~(\ref{eq:overlapcom}), (\ref{eq:tcom}), 
 and (\ref{eq:vcom}) provide the overlap and Hamiltonian matrices in 
 Eqs.~(\ref{eq:scom}), thus the eigenenergies
 and eigenfunctions of the uncoupled c.m.\ motion can be found.

 \subsubsection{Matrix elements of the coupled Hamiltonian}

 Finally, for obtaining the coupled solutions the Hamiltonian matrix 
 elements $H_{\kappa,\kappa'}$ in Eq.~(\ref{eq:Hkk}) have to be 
 calculated. Remind, the total Hamiltonian $\oper{H}$ was written as 
 a sum of the uncoupled Hamiltonians of rel.\ and c.m.\ motion, 
 $\oper{h}_{\rm rel.}, \oper{h}_{\rm c.m.}$, and the  
 coupling term $\oper{W}$ (\ref{eq:hamcmrelspfr}). Since the 
 configurations are build with the eigenfunctions $\varphi$ and 
 $\psi$ of the uncoupled Hamiltonians, only the simple diagonal 
 contribution 
 \begin{align}
   \ds \opij{\Phi_\kappa}{\oper{h}_{\rm c.m.}
                    &+\oper{h}_{\rm rel.}}{\Phi_{\kappa'}} 
       {}
       \nonumber\\
       {}
       &=
       \opij{\varphi_{i_\kappa}\psi_{j_\kappa}}{\oper{h}_{\rm c.m.}
              +\oper{h}_{\rm
                    rel.}}{\varphi_{i_{\kappa'}}\psi_{j_{\kappa'}}}
       {}
       \nonumber\\
       {}
       &= ( \epsilon_{i_\kappa}^{\rm rel.} 
           + \epsilon_{j_\kappa}^{\rm c.m.} )
	   \,\delta_{i_{\kappa\vphantom{'}},i_{\kappa'}} 
	   \,\delta_{j_{\kappa\vphantom{'}},j_{\kappa'}} 
 \end{align}
 is obtained from $\oper{h}_{\rm rel.}$ and $\oper{h}_{\rm c.m.}$. 

 The remaining task is thus the calculation of the matrix elements 
 that couple rel.\ and c.m.\ motions, i.\,e., the ones of $\oper{W}$. 
 They are given as
 \begin{widetext}
 \begin{multline}
     \ds
         W_{\kappa,\kappa'} =
             \frac12 \sum\limits_{s=1}^2
             \sum\limits_{c=x,y,z}
             V^s_c \Biggl\{(-1)^{\eta_s}
             \sum\limits_{j=0}^{n-1}
             (-1)^j\frac{(2k_c\mu_{\eta_s})^{2j+1}}
             {(2j+1)!}
             \sum\limits_{\alpha=2}^{N_r-1}
             \sum\limits_{l=0}^{N_l}
             \sum\limits_{m=-l}^{l}
             \tilde{c}_{p_\kappa,\mathbf{a}}^{\rm rel.}
             \sum\limits_{\alpha'=2}^{N_r-1}
             \sum\limits_{l'=0}^{N_l}
             \sum\limits_{m'=-l}^{l}
             \tilde{c}_{p_{\kappa'},\mathbf{a}'}^{\rm rel.}
	     \,\mathbb{B}^{2j+1}_{\alpha\, \alpha'}
	      \\
	      \ds
             \times\, 
             \sum\limits_{l_j=1,\{2\}}^{2j+1}
             \Biggl[\Biggl(\tilde{\mathbb{Y}}_{l_j0j}^c
             \begin{pmatrix} l_j & l & l' \\
             0 & m & m' \end{pmatrix}
	     +\,
             \sum\limits_{m_j=-l_j,\{2\}}^{l_j}
             \tilde{\mathbb{Y}}_{l_jm_jj}^c
             \begin{pmatrix} l_j & l & l' \\
             m_j & m & -m' \end{pmatrix}\Biggr)
              \mathbb{A}^{m'}_{l_jll'}
             \begin{pmatrix} l_j & l & l' \\
             0 & 0 & 0 \end{pmatrix}
             \Biggr]
	     {}
              \\
	     {}
	     \ds
	     \times\,
	     \sum\limits_{i=0}^{n-1-j}
             (-1)^i\frac{(2k_c)^{2i+1}}
             {(2i+1)!}
             \sum\limits_{\beta=2}^{N_R-1}
             \sum\limits_{L=0}^{N_L}
             \sum\limits_{M=-L}^{L}
             \tilde{c}_{q_\kappa,\mathbf{b}}^{\rm c.m.}
             \sum\limits_{\beta'=2}^{N_R-1}
             \sum\limits_{L'=0}^{N_L}
             \sum\limits_{M'=-L'}^{L'}
             \tilde{c}_{q_{\kappa'},\mathbf{b}'}^{\rm c.m.}
	     \,\mathbb{B}^{2i+1}_{\beta\, \beta'}
	     {}
              \\
	     {}
	     \ds
             \times\,
	     \sum\limits_{L_i=1,\{2\}}^{2i+1}
             \Biggl[
	       \Biggl(\tilde{\mathbb{Y}}_{L_i0i}^c
             \begin{pmatrix} L_i & L & L' \\
             0 & M & M' \end{pmatrix}   
	     +
             \sum\limits_{M_i=-L_i,\{2\}}^{L_i}
             \tilde{\mathbb{Y}}_{L_iM_ii}^c
             \begin{pmatrix} L_i & L & L' \\
             M_i & M & -M' \end{pmatrix} 
             \Biggr) \mathbb{A}^{M'}_{L_iLL'}
              \begin{pmatrix} L_i & L & L' \\
              0 & 0 & 0 \end{pmatrix} \Biggr] 
	     {}
               \\
	     {}
	     \ds
             -\, \sum\limits_{t=1}^{n}
             (-1)^t\frac{(2k_c\mu_{\eta_s})^{2t}}{(2t)!}
             \sum\limits_{\alpha=2}^{N_r-1}
             \sum\limits_{l=0}^{N_l}
             \sum\limits_{m=-l}^{l}
             \tilde{c}_{p_\kappa,\mathbf{a}}^{\rm rel.}
	     \sum\limits_{\alpha'=2}^{N_r-1}
             \sum\limits_{l'=0}^{N_l}
             \sum\limits_{m'=-l}^{l}
             \tilde{c}_{p_{\kappa'},\mathbf{a}'}^{\rm rel.}
	     \,\mathbb{B}^{2t}_{\alpha\, \alpha'}
	     {}
              \\
	     {}
	     \ds
             \times\, \sum\limits_{l_t=0,\{2\}}^{2t}
             \sum\limits_{m_t=-l_t,\{2\}}^{l_t}
             \mathbb{Y}_{l_tm_tt}^c
             \mathbb{A}^{m'}_{l_tll'}
             \begin{pmatrix} l_t & l & l' \\
             m_t & m & -m' \end{pmatrix}
             \begin{pmatrix} l_t & l & l' \\
             0 & 0 & 0 \end{pmatrix}
             \sum\limits_{k=1}^{n-t}
             (-1)^k\frac{(2k_c)^{2k}}{(2k)!}
             \sum\limits_{\beta=2}^{N_R-1}
             \sum\limits_{L=0}^{N_L}
             \sum\limits_{M=-L}^{L}
             \tilde{c}_{q_\kappa,\mathbf{b}}^{\rm c.m.}
	     {}
	      \\
	     {}
	     \ds
             \times\, \sum\limits_{\beta'=2}^{N_R-1} 
             \sum\limits_{L'=0}^{N_L}
             \sum\limits_{M'=-L'}^{L'}
             \tilde{c}_{q_{\kappa'},\mathbf{b}'}^{\rm c.m.}
             \,\mathbb{B}^{2k}_{\beta\, \beta'}
	     \sum\limits_{L_k=0,\{2\}}^{2k}
             \sum\limits_{M_k=-L_k,\{2\}}^{L_k}
             \mathbb{Y}_{L_kM_kk}^c
             \mathbb{A}^{M'}_{L_kLL'}
             \begin{pmatrix} L_k & L & L' \\
             M_k & M & -M' \end{pmatrix}
             \begin{pmatrix} L_k & L & L' \\
             0 & 0 & 0 \end{pmatrix}
             \Biggr\}\, .
	     \label{eq:wij}
   \end{multline}
\end{widetext}

 Despite the fact that Eq.~(\ref{eq:wij}) is somewhat lengthy it is
 convenient and practical for computational purposes. 
 While in the computer implementation the summations are 
 ordered in such a fashion that the numerical efforts are minimized, 
 the order given in Eq.~(\ref{eq:wij}) is more transparent.

 \section{Symmetry of the system}
 \label{sec:symofsys}

\begin{figure}[!ht]
  \centering
  \includegraphics[width=0.50\textwidth]{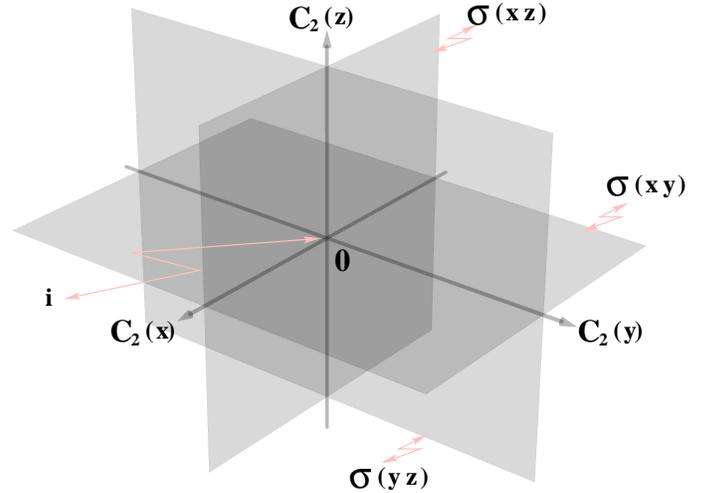}
  \caption[The symmetry elements of the two particles in a $\sin^2$
           trap] 
      {{\footnotesize
      The symmetry elements of the two particles
      interacting by a central potential in a
      $\sin^2$-like trap. The list is complete
      with the identity element E added.
    } }
  \label{fig:d2hsymelem}
 \end{figure}

 The Hamiltonian of two atoms interacting via a central potential 
 and trapped in $\sin^2$-like or $\cos^2$-like potentials that are 
 oriented along three orthogonal directions is 
 invariant under the symmetry operations of the point group 
 D$_{\rm 2h}$. Since the optical-lattice potential is chosen 
 along the three Cartesian axes $x$, $y$, and $z$, the single particle
 Hamiltonians in Eq.~(\ref{eq:single_part_hamil})    
 $\ds
          \oper{\mathcal{H}}_j(x,y,z)    = \oper{\mathcal{H}}_j(-x,-y,-z) = 
          \oper{\mathcal{H}}_j(-x,y,z)  = \oper{\mathcal{H}}_j(x,-y,z)   =
          \oper{\mathcal{H}}_j(x,y,-z)  = \oper{\mathcal{H}}_j(-x,-y,z)  = 
          \oper{\mathcal{H}}_j(x,-y,-z) = \oper{\mathcal{H}}_j(-x,y,-z)$
 are equivalent. This is a consequence of the symmetry elements 
 of the orthorhombic D$_{\rm 2h}$ group that contains besides the 
 identity operation $E$ and the inversion symmetry {\bf i} also three 
 twofold rotations (by the angle $\pi$) $C_2 (x)$, $C_2 (y)$,
 and $C_2 (z)$ as well as the three mirror planes $\sigma(xy)$, 
 $\sigma (xz)$, and $\sigma (yz)$. The symmetry elements 
 are illustrated in Fig.~\ref{fig:d2hsymelem}. 

 The symmetry group D$_{\rm 2h}$ has eight irreducible
 representations (see Table~\ref{tab:d2hchartab}): $A_g$, $B_{1g}$, 
 $B_{2g}$, $B_{3g}$, $A_u$, $B_{1u}$, $B_{2u}$, $B_{3u}$. Clearly, 
 the explicit use of  
 symmetry is advantageous, since it reduces the numerical efforts 
 dramatically as the different irreducible representations can be 
 treated independently of each other. This reduces the dimensions 
 of the matrices that have to be diagonalized by approximately a 
 factor of $8\times 8=64$. Furthermore, many integrals vanish  
 for symmetry reasons and have thus not to be calculated at all.
 \begin{table}
    \caption{Character table of the D$_{\rm 2h}$ point group}
    \begin{ruledtabular}
     \begin{tabular}{c|cccccccc}
         \vspace{-0.3cm}
         &     &    &     &     &    &    &    &    \\ 
         \vspace{-0.3cm}
         D$_{2{\rm h}}$ & $E$ & $C_2(z)$ & $C_2(y)$ & $C_2(x)$ & 
         {\bf i} & $\sigma(xy)$ & $\sigma(xz)$ & $\sigma(yz)$
         \\      
         &     &    &     &     &    &    &    &    \\  
       \hline
         \vspace{-0.3cm}
              &     &    &     &     &    &    &    &    \\ 
         \vspace{-0.3cm}
 \,$A_g$\,    &  \,1\,   &  \,1\,  &  \,1\,   &  \,1\,   &  \,1\,  
              & \,1\,   &  \,1\,  &  \,1\,  \\  
              &     &    &     &     &    &    &    &    \\  
         \vspace{-0.3cm}
              &     &    &     &     &    &    &    &    \\ 
          \vspace{-0.3cm}
 \,$B_{1g}$\, &  \,1\,   &  \,1\,  &  \,-1\,   &  \,-1\,   &  \,1\,  
              & \,1\,   &  \,-1\,  &  \,-1\,  \\  
              &     &    &     &     &    &    &    &    \\  
         \vspace{-0.3cm}
              &     &    &     &     &    &    &    &    \\  
         \vspace{-0.3cm}
 \,$B_{2g}$\, &  \,1\,   &  \,-1\,  &  \,1\,   &  \,-1\,   &  \,1\,  
              & \,-1\,   &  \,1\,  &  \,-1\,  \\  
              &     &    &     &     &    &    &    &    \\  
         \vspace{-0.3cm}
              &     &    &     &     &    &    &    &    \\  
         \vspace{-0.3cm}
 \,$B_{3g}$\, &  \,1\,   &  \,-1\,  &  \,-1\,   &  \,1\,   &  \,1\,  
              & \,-1\,   &  \,-1\,  &  \,1\,  \\  
              &     &    &     &     &    &    &    &    \\  
         \vspace{-0.3cm}
              &     &    &     &     &    &    &    &    \\ 
          \vspace{-0.3cm}
 \,$A_u$\,    &  \,1\,   &  \,1\,  &  \,1\,   &  \,1\,   &  \,-1\,  
              & \,-1\,   &  \,-1\,  &  \,-1\,  \\  
              &     &    &     &     &    &    &    &    \\  
         \vspace{-0.3cm}
              &     &    &     &     &    &    &    &    \\  
         \vspace{-0.3cm}
 \,$B_{1u}$\, &  \,1\,   &  \,1\,  &  \,-1\,   &  \,-1\,   &  \,-1\,  
              & \,-1\,   &  \,1\,  &  \,1\,  \\  
              &     &    &     &     &    &    &    &    \\  
         \vspace{-0.3cm}
              &     &    &     &     &    &    &    &    \\ 
         \vspace{-0.3cm}
 \,$B_{2u}$\, &  \,1\,   &  \,-1\,  &  \,1\,   &  \,-1\,   &  \,-1\,  
              & \,1\,   &  \,-1\,  &  \,1\,  \\  
              &     &    &     &     &    &    &    &    \\  
         \vspace{-0.3cm}
              &     &    &     &     &    &    &    &    \\  
         \vspace{-0.3cm}
 \,$B_{3u}$\, &  \,1\,   &  \,-1\,  &  \,-1\,   &  \,1\,   &  \,-1\,  
              & \,1\,   &  \,1\,  &  \,-1\,  \\  
              &     &    &     &     &    &    &    &    \\  
    \end{tabular}
    \end{ruledtabular}
   \label{tab:d2hchartab}
 \end{table}

 In fact, the D$_{\rm 2h}$ symmetry is a consequence of the considered 
 shape of the potential and thus the symmetry of the single-particle 
 Hamiltonians in absolute Cartesian coordinates, $\oper{\mathcal{H}}_j$ in 
 Eq.~(\ref{eq:single_part_hamil}). Since the 
 atom-atom interaction $\oper{u}$ is invariant under all operations 
 in D$_{\rm 2h}$, the total Hamiltonian $\oper{\mathcal{H}}$ in Eq.~(\ref{eq:origham}) belongs also to the 
 D$_{\rm 2h}$ group. As may be less transparent on a first glance, but 
 can be shown from a complete symmetry analysis,      
 also the rel.\ and c.m.\ Hamiltonians 
 $\oper{h}_{\rm rel.}(\vec{r})$ and $\oper{h}_{\rm c.m.}(\vec{R})$
 possess the same symmetry as the total Hamiltonian. Therefore, it is
 sufficient to examine the symmetry properties, e.\,g., for the
 rel.\ part only. The c.m.\ part has the same properties and the ones of
 the total Hamiltonian can then be deduced from the properties of the 
 direct tensor products. 

 \begin{table}
    \caption[Results of the D$_{\rm 2h}$ group operations]
             {Results of the D$_{\rm 2h}$ group operations
             on the absolute and spherical coordinates, and 
             the corresponding transformations of spherical 
             harmonics. Given are the values of $a,\,b,\,c$
             that are multipliers for the
             $x,\, y,\, z$ coordinates, respectively, and 
             $\theta',\,\phi'$ that are shifts of the spherical
             coordinates $\theta,\,\phi$, respectively.}
   \begin{ruledtabular}
     \begin{tabular}{c|ccc}
    \vspace{-0.3cm}
            &                     &                         & \\     
      \vspace{-0.3cm}
        \text{symmetry}    &\; Absolute \; &\;  Spherical  \;  
                           & \;  $Y_l^m$  \; \\
    \vspace{-0.04cm}
            &                     &                         & \\     
      \vspace{-0.3cm}
            &$(a\,x,b\,y,c\,z)$ & ($\theta'+\theta$,\,$\phi'+\phi$) 
            & $Y_l^m(\theta^{\prime}+\theta,\phi^{\prime}+\phi)$ \\
            &                     &                         &  \\
\hline
    \vspace{-0.3cm}
            &                      &                         &  \\
    \vspace{-0.3cm}
\;E\;
            &  $(1,1,1)$ &  ($0+\theta,0+\phi$)    
            & $Y_l^m(\theta,\phi)$ \\
            &                      &                         &   \\
    \vspace{-0.3cm}
            &                      &                          &   \\
    \vspace{-0.3cm}
\;$C_2(z)$\;
            &$(-1,-1,1)$  & $(0+\theta,\pi+\phi)$   
            &$(-1)^{m}\,Y_l^m(\theta,\phi)$   \\
            &                      &                          &   \\
    \vspace{-0.3cm}
            &                      &                          &   \\
    \vspace{-0.3cm}
\;$C_2(y)$\;
            &$(-1,1,-1)$  & $(\pi-\theta,\pi-\phi)$   
            &$(-1)^{l+m}\,Y_l^{-m}(\theta,\phi)$   \\
            &                      &                          &   \\
    \vspace{-0.3cm}
            &                      &                          &   \\
    \vspace{-0.3cm}
\;$C_2(x)$\;
            &$(1,-1,-1)$  & $(\pi-\theta,2\pi-\phi)$   
            &$(-1)^l\,Y_l^{-m}(\theta,\phi)$   \\
            &                      &                          &   \\
    \vspace{-0.3cm}
            &                      &                          &   \\
    \vspace{-0.3cm}
\;{\bf i}\;
            &$(-1,-1,-1)$&  ($\pi-\theta,\pi+\phi$) 
            &$(-1)^l\,Y_l^m(\theta,\phi)$    \\
            &                      &                          &  \\
    \vspace{-0.3cm}
            &                      &                          &   \\
    \vspace{-0.3cm}
\;$\sigma(xy)$\;
            &$(1,1,-1)$  & $(\pi-\theta,0+\phi)$   
            &$(-1)^{l+m}\,Y_l^m(\theta,\phi)$   \\
            &                      &                          &   \\
    \vspace{-0.3cm}
            &                      &                          &   \\
    \vspace{-0.3cm}
\;$\sigma(xz)$\;
            &$(1,-1,1)$  & $(0+\theta,2\pi-\phi)$   
            &$(-1)^{m}\,Y_l^{-m}(\theta,\phi)$   \\
            &                      &                          &   \\
    \vspace{-0.3cm}
            &                      &                          &   \\
    \vspace{-0.3cm}
\;$\sigma(yz)$\;
            &$(-1,1,1)$  &($0+\theta,\pi-\phi$)      
            &$Y_l^{-m}(\theta,\phi)$  
\\
            &                      &                          &   \\
    \end{tabular}
    \end{ruledtabular}
   \label{tab:acscylm}
 \end{table}
 In order to use the symmetry when solving the eigenvalue problems 
 of the uncoupled rel.\ and c.m.\ motions, symmetry-adapted basis 
 functions have to be obtained. Since all basis functions adopted 
 in this work are centered at the origin and are products of a radial 
 part times a spherical harmonic, (\ref{eq:wfrel2}) 
 and (\ref{eq:wfcom2}), the symmetry operations affect only the angular 
 part. Therefore, linear combinations of the spherical harmonics have 
 to be found that transform like the irreducible representations of 
 the D$_{\rm 2h}$ group. The problem of determining symmetry-adapted 
 basis functions from the complete set of spherical harmonics has a 
 long history, starting with the introduction of cubic harmonics 
 for the cubic point group~\cite{gen:lage47}. Since it appears, however, 
 to be not that trivial to find the orthorhombic harmonics in easily 
 accessible form, they are given explicitly together with a brief 
 derivation. First, the action of the symmetry elements of D$_{\rm 2h}$ 
 on the spherical harmonics has to be considered. The result is 
 shown in Table~\ref{tab:acscylm} that provides also the intermediate 
 steps, the result of applying the symmetry operations on the 
 Cartesian and the spherical coordinates. The most important result 
 is that all symmetry operations of D$_{\rm 2h}$ leave $l$ unchanged, 
 i.\,e., only $Y_l^m$ with identical values of $l$ are transformed into 
 each other. As a consequence, the symmetry-adapted basis functions 
 are superpositions of spherical harmonics with a fixed value of $l$. 
 
 In view of the eight irreducible representations of D$_{\rm 2h}$ 
 (see Table~\ref{tab:d2hchartab}) one needs to find the required eight
 sets of orthonormal linear combinations of spherical harmonics. 
 In the present case, this is easily achieved using the standard 
 projector technique, i.\,e., by applying the projector 
 \begin{equation}
   \oper{P}_i = \frac{1}{h} \sum_{k=1}^{h} 
                        \: \chi_i(\oper{O}_k)^*\, \oper{O}_k
 \label{eq:Projector}
 \end{equation}
 of the irreducible representation $i$ onto a spherical harmonic 
 $Y_l^m$. In Eq.~(\ref{eq:Projector}) it is used that all 
 irreducible representations in D$_{\rm 2h}$ are non-degenerate. 
 Furthermore, $h$ is the total number of symmetry 
 operations (eight for D$_{\rm 2h}$), $\oper{O}_k$ the operator
 corresponding to symmetry element $k$, and $\chi_i(\oper{O}_k)$ the
 character of symmetry element $k$ for the irreducible representation
 $i$. While all characters are listed in Table \ref{tab:d2hchartab}, 
 the results of the symmetry operations on $Y_l^m$ are given 
 in the last column of Table~\ref{tab:acscylm}. For example, 
 the application of $\oper{P}_{B_{3u}}$ gives 
 \begin{multline}
  \oper{P}_{B_{3u}} \, Y_l^m = \frac{1}{8} \:
          \left[\, 1 - (-1)^m - (-1)^l + (-1)^{l+m} \,\right]  \:
          Y_l^m
	  {}
	  \\
	  {}
           \: + \: 
          \left[\, (-1)^l - (-1)^{l+m} + (-1)^m - 1\,\right] \: Y_l^{-m} \,.
 \label{eq:B3u}
 \end{multline}
 Clearly, only the combination of odd values of $l$ and $m$ yields in 
 this case a non-zero result and thus a symmetry-adapted basis function, 
 \begin{equation}
  \oper{P}_{B_{3u}} \, Y_l^m 
            = \frac{1}{2} \: \left( \, Y_l^m - Y_l^{-m} \, \right) 
             \qquad l,m \quad {\rm odd}\, .
 \label{eq:B3u_final}
 \end{equation}
 The use of these symmetry-adapted basis functions (superposition 
 of spherical harmonics instead of a single one and restriction 
 on $l$ and $m$) modifies the wave functions of the rel.\ motion 
 in Eq.~(\ref{eq:wfrel2}) into
 \begin{eqnarray}
       \varphi_i^{A_g} &=& 
                 \sum\limits_{\alpha=1}^{N_r}
                 \sum\limits_{l=0,\{2\}}^{N_l}
                 \sum\limits_{m=0,\{2\}}^l
                 \tilde{c}_{i, \alpha l m}^{A_g} \,r^{-1}\,B_{\alpha} (r)
                 \mathscr{Y}_{lm}^+ 
                \\ 
       \varphi_i^{B_{1g}} &=& 
                 \sum\limits_{\alpha=1}^{N_r}
                 \sum\limits_{l=2,\{2\}}^{N_l}
                 \sum\limits_{m=2,\{2\}}^l
                 \tilde{c}_{i, \alpha l m}^{B_{1g}} \,r^{-1}\,B_{\alpha} (r)
                 \mathscr{Y}_{lm}^+ 
                \\ 
       \varphi_i^{B_{2g}} &=& 
                 \sum\limits_{\alpha=1}^{N_r}
                 \sum\limits_{l=2,\{2\}}^{N_l}
                 \sum\limits_{m=1,\{2\}}^l
                 \tilde{c}_{i, \alpha l m}^{B_{2g}} \,r^{-1}\,B_{\alpha} (r)
                 \mathscr{Y}_{lm}^- 
                \\ 
       \varphi_i^{B_{3g}} &=& 
                 \sum\limits_{\alpha=1}^{N_r}
                 \sum\limits_{l=2,\{2\}}^{N_l}
                 \sum\limits_{m=1,\{2\}}^l
                 \tilde{c}_{i, \alpha l m}^{B_{3g}} \,r^{-1}\,B_{\alpha} (r)
                 \mathscr{Y}_{lm}^+ 
                \\ 
       \varphi_i^{A_{u}} &=& 
                 \sum\limits_{\alpha=1}^{N_r}
                 \sum\limits_{l=3,\{2\}}^{N_l}
                 \sum\limits_{m=2,\{2\}}^l
                 \tilde{c}_{i, \alpha l m}^{A_{u}} \,r^{-1}\,B_{\alpha} (r)
                 \mathscr{Y}_{lm}^- 
                \\ 
       \varphi_i^{B_{1u}} &=& 
                 \sum\limits_{\alpha=1}^{N_r}
                 \sum\limits_{l=1,\{2\}}^{N_l}
                 \sum\limits_{m=0,\{2\}}^l
                 \tilde{c}_{i, \alpha l m}^{B_{1u}} \,r^{-1}\,B_{\alpha} (r)
                 \mathscr{Y}_{lm}^+ 
                \\ 
       \varphi_i^{B_{2u}} &=& 
                 \sum\limits_{\alpha=1}^{N_r}
                 \sum\limits_{l=1,\{2\}}^{N_l}
                 \sum\limits_{m=1,\{2\}}^l
                 \tilde{c}_{i, \alpha l m}^{B_{2u}} \,r^{-1}\,B_{\alpha} (r)
                 \mathscr{Y}_{lm}^+ 
                \\ 
       \varphi_i^{B_{3u}} &=& 
                 \sum\limits_{\alpha=1}^{N_r}
                 \sum\limits_{l=1,\{2\}}^{N_l}
                 \sum\limits_{m=1,\{2\}}^l
                 \tilde{c}_{i, \alpha l m}^{B_{3u}} \,r^{-1}\,B_{\alpha} (r)
                 \mathscr{Y}_{lm}^- 
 \end{eqnarray}
 where 
 \begin{eqnarray}
   \mathscr{Y}_{l0}^{+} &=& \mathscr{Y}_{l0}^{-} = Y_l^0
              (\theta,\phi)\, , 
              \\
   \mathscr{Y}_{lm}^{\pm} &=& 
            Y_l^m(\theta,\phi)\pm Y_l^{-m}(\theta,\phi) \quad ( m\neq 0 )
 \end{eqnarray}
 is introduced for compactness and a summation index $l=i,\{2\}$ means $l=i,
 i+2, \dots$ Moreover, for the coefficients
 $\tilde{c}$ the superscripts rel. are omitted for better readability.  Clearly, the
 consideration of symmetry-adapted basis functions for the c.m.\ motion
 leads to a completely analogous modification of Eq.~(\ref{eq:wfcom2}).

 \begin{table}
    \caption{Product table of the D$_{\rm 2h}$ point group}
    \begin{ruledtabular}
     \begin{tabular}{c|cccccccc}
       \vspace{-0.3cm}
              &     &    &     &     &    &    &    &    \\  
       \vspace{-0.3cm}
              $\otimes$&  \;$A_g$\; &\;$B_{1g}$\; & \;$B_{2g}$\;&  
              \;$B_{3g}$\; & \;$A_u$\;  &\;$B_{1u}$\; & 
              \;$B_{2u}$\;  & \;$B_{3u}$\;  \\
              &     &    &     &     &    &    &    &    \\
       \hline
       \vspace{-0.3cm}
              &     &    &     &     &    &    &    &    \\
       \vspace{-0.3cm}
 \;$A_g$\;    &  \;$A_g$\; & \;$B_{1g}$\; & \;$B_{2g}$\;
              &  \;$B_{3g}$\;   &  \;$A_u$\;  & \;$B_{1u}$\;
              &  \;$B_{2u}$\;  &  \;$B_{3u}$\;  \\
              &     &    &     &     &    &    &    &    \\
       \vspace{-0.3cm}
              &     &    &     &     &    &    &    &    \\
       \vspace{-0.3cm}
 \;$B_{1g}$\; &  \;$B_{1g}$\;   &  \;$A_g$\;  & \;$B_{3g}$\;
              & \;$B_{2g}$\;   &  \;$B_{1u}$ \;  & \;$A_u$\;
              &  \;$B_{3u}$\;  &  \;$B_{2u}$\;
              \\
              &     &    &     &     &    &    &    &    \\
       \vspace{-0.3cm}
              &     &    &     &     &    &    &    &    \\
       \vspace{-0.3cm}
 \;$B_{2g}$\; &  \;$B_{2g}$\;   &  \;$B_{3g}$\;  & \;$A_g$\;
              & \;$B_{1g}$\;   &  \;$B_{2u}$ \;  & \;$B_{3u}$\;   
              &  \;$A_u$\;  &  \;$B_{1u}$\;
              \\   
              &     &    &     &     &    &    &    &    \\  
       \vspace{-0.3cm}
              &     &    &     &     &    &    &    &    \\
         \vspace{-0.3cm}
 \;$B_{3g}$\; &  \;$B_{3g}$\;   &  \;$B_{2g}$\;  & \;$B_{1g}$\;    
              & \;$A_g$\;   &  \;$B_{3u}$ \;  & \;$B_{2u}$\;   
              &  \;$B_{1u}$\;  &  \;$A_u$\;
              \\   
              &     &    &     &     &    &    &    &    \\  
       \vspace{-0.3cm}
              &     &    &     &     &    &    &    &    \\ 
        \vspace{-0.3cm}
 \;$A_u$\; &  \;$A_u$\;   &  \;$B_{1u}$\;  & \;$B_{2u}$\;
              & \;$B_{3u}$\;   &  \;$A_g$ \;  & \;$B_{1g}$\;   
              &  \;$B_{2g}$\;  &  \;$B_{3g}$\;
              \\   
              &     &    &     &     &    &    &    &    \\  
       \vspace{-0.3cm}
              &     &    &     &     &    &    &    &    \\ 
        \vspace{-0.3cm}
 \;$B_{1u}$\; &  \;$B_{1u}$\;   &  \;$A_u$\;  & \;$B_{3u}$\;
              & \;$B_{2u}$\;   &  \;$B_{1g}$ \;  & \;$A_g$\;   
              &  \;$B_{3g}$\;  &  \;$B_{2g}$\;
              \\   
              &     &    &     &     &    &    &    &    \\ 
       \vspace{-0.3cm}
              &     &    &     &     &    &    &    &    \\ 
        \vspace{-0.3cm}
 \;$B_{2u}$\; &  \;$B_{2u}$\;   &  \;$B_{3u}$\;  & \;$A_u$\;
              & \;$B_{1u}$\;   &  \;$B_{2g}$ \;  & \;$B_{3g}$\;   
              &  \;$A_g$\;  &  \;$B_{1g}$\;
              \\   
              &     &    &     &     &    &    &    &    \\ 
       \vspace{-0.3cm}
              &     &    &     &     &    &    &    &    \\  
       \vspace{-0.3cm}
 \;$B_{3u}$\; &  \;$B_{3u}$\;   &  \;$B_{2u}$\;  & \;$B_{1u}$\;
              & \;$A_u$\;   &  \;$B_{3g}$ \;  & \;$B_{2g}$\;   
              &  \;$B_{1g}$\;  &  \;$A_g$\;
              \\   
              &     &    &     &     &    &    &    &    \\ 
    \end{tabular}
    \end{ruledtabular}
   \label{tab:d2hprodtab}
 \end{table}
 Since the D$_{\rm 2h}$ point group contains only non-degenerate 
 irreducible representations, its product table (showing the 
 result of a direct tensor product between a pair of irreducible 
 representations and given in Tab.~\ref{tab:d2hprodtab}) is 
 straightforwardly obtained and every product corresponds 
 uniquely to one irreducible representation. For example, a 
 configuration $\Phi_\kappa=\varphi_{i_\kappa}^{B_{3g}} \, 
 \psi_{j_\kappa}^{B_{2g}}$ transforms as $B_{1g}$. Clearly, 
 symmetry-adapted configurations are straightforwardly constructed 
 from the symmetry-adapted rel.\ and c.m.\ orbitals. 

 In the case of indistinguishable atoms, the quantum statistics 
 has to be considered. For Fermionic atoms, the total
 wavefunction 
 must change sign under particle exchange, while it must remain 
 the same for Bosons. Particle exchange does not influence the 
 c.m.\ coordinate (i.\,e., $\vec{R} \rightarrow \vec{R}$ or, equivalently,  
 $\Phi \rightarrow \Phi,\, \Theta \rightarrow \Theta$), but the
 rel.\ coordinate (i.\,e., $\vec{r} \rightarrow -\vec{r}$ or 
 $\phi \rightarrow \pi+\phi,\, \theta \rightarrow \pi-\theta$). 
 Since particle exchange corresponds to the symmetry operation 
 of inversion ({\bf i}) for the \rm coordinate, all {\it gerade} 
 basis functions ($\varphi_{A_g}$, $\varphi_{B_{1g}}$, 
 $\varphi_{B_{2g}}$, and $\varphi_{B_{3g}}$) are allowed for 
 identical Bosons, the {\it ungerade} functions 
 ($\varphi_{A_u}$, $\varphi_{B_{1u}}$, $\varphi_{B_{2u}}$, and 
 $\varphi_{B_{3u}}$) for identical Fermions. The quantum statistics 
 for indistinguishable atoms is thus easily taken into account 
 and reduces the number of possible orbital combinations by 
 factor 2. The straightforward (almost automatic) implementation 
 of the quantum statistics that leads even to a direct reduction 
 of the computational demands is a further advantage of the 
 present approach.

\section{Computational details}
\label{sec:compdet}

 The theoretical approach presented in this work provides an 
 efficient way to treat two interacting atoms in an orthorhombic 
 optical lattice. The use of c.m.\ and rel.\ coordinates and the 
 expansion of the basis functions in $B$ splines for the radial 
 part and spherical harmonics for the angular parts is especially 
 useful for considering realistic interatomic (molecular) interaction 
 potentials. Furthermore, the Bosonic or Fermionic nature of the atoms 
 is easily accounted for. However, in the case of strongly anisotropic 
 lattice potentials the advantage of the use of spherical harmonics 
 that all involved integrals can be efficiently and analytically 
 calculated is partly compensated by their slow convergence. Similarly, 
 the adopted exact diagonalization approach for incorporating the 
 coupling of c.m.\ and rel.\ motion has the advantage of being exact, 
 if converged, but is known to be slowly convergent. For many 
 experimentally relevant parameters the calculations are, therefore,
 still very demanding and thus an efficient implementation of all 
 involved computational steps was mandatory. The adequate choice 
 of the basis-set parameters adapted to the considered problem 
 can improve the efficiency drastically. Thus it is worthwhile 
 to at least briefly discuss some technical aspects of both 
 the implementation and the choice of basis sets.

\subsection{Interatomic interaction potential}
\label{subsec:interaction}

 In the present approach the interatomic interaction enters the 
 calculation only in the determination of the rel.\ orbitals. 
 In the so far considered case of isotropic interaction potentials 
 the potential influences only the calculation of the radial 
 integral~(\ref{eq:urel}). Clearly, an extension to orientation-dependent 
 interaction potentials (like dipole-dipole interaction) is possible 
 by expanding the angular part of the interaction potential in 
 spherical harmonics. Then the resulting angular integrals can still 
 be solved analytically. Since the radial integral~(\ref{eq:urel})
 is calculated using quadrature, even a usually only numerically 
 given Born-Oppenheimer potential curve can be used in order to 
 achieve a realistic description of the interatomic interaction. 
 Clearly, any type of potential can be easily used. Only the 
 implementation of the $\delta$-function pseudopotential required 
 some care due to its numerically ill-behaved nature.

 In the ultracold regime the scattering process is extremely 
 sensitive to all details of the complete interatomic potential 
 curve. In fact, for most experimentally relevant alkali atoms, 
 it is impossible to calculate the potential curves with sufficient 
 precision. In the zero-energy limit the scattering process is 
 fully described by the scattering length $a_{\rm sc}$ 
 \cite{cold:wein99}. Depending on the considered atoms, their 
 isotopes, and electronic states $a_{\rm sc}$ can have 
 very different values for different systems, 
 ranging from $a_{\rm sc}\gg 0$ (strongly repulsive) via 
 $a_{\rm sc}\approx 0$ (almost non-interacting) to 
 $a_{\rm sc}\ll 0$ (strongly attractive). An important aspect 
 of ultracold atomic gases is the fact that many atom pairs 
 possess magnetic Feshbach resonances, i.\,e., with a magnetic 
 field the colliding atoms can be brought into resonance with 
 a molecular bound state. As a consequence, the scattering length 
 becomes experimentally tunable \cite{cold:loft02,cold:rega03b}. 
 This method was also successfully used to tune the interatomic 
 interaction between atoms in the optical lattices, 
 see, e.\,g., \cite{cold:koeh05,cold:stoe06,cold:ospe06a}. 

 However, the correct theoretical treatment of a magnetic Feshbach 
 resonance requires the solution of a multichannel problem that 
 is numerically demanding due to the different length scales 
 involved. Within an optical lattice the influence of the 
 confining potential on the multichannel problem has also to 
 be properly considered (see \cite{cold:schn11} and references 
 therein). To perform such a study within the present algorithm 
 appears prohibitively difficult, at least with the computer 
 resources at our disposal. On the other hand, the variation 
 of the interaction strength (characterized in the trap-free 
 situation and at the zero-energy limit by $a_{\rm sc}$) can 
 be mimicked within single-channel models. In this case, 
 some parameter in the rel.\ motion Hamiltonian is varied 
 in such a fashion that resonant behavior occurs. Whenever 
 the manipulation leads to a shift of a very weakly bound state 
 into the dissociation continuum, resonant behavior (divergence 
 of $a_{\rm sc}$) is observed. Examples for such artificially 
 obtained single-channel resonances include the variation of 
 van der Waals coefficients~\cite{cold:ribe04}, the reduced mass 
 \cite{cold:gris07}, the inner-wall of the molecular interaction 
 potential \cite{cold:gris09}, or the local modification of the 
 Born-Oppenheimer curve at intermediate distances 
 \cite{cold:gris10}. A comparison of these different procedures 
 and the full multi-channel treatment is provided in \cite{cold:gris10}. 
 As is shown in \cite{cold:schn09} a better and in fact almost 
 perfect model for a multi-channel Feshbach resonance can be 
 obtained with a two-channel model which appears more realistic 
 with respect to a possible implementation within the present 
 approach than the full multi-channel Hamiltonian.

\subsection{Basis-set considerations}
\label{subsec:basisset}

 Due to the choice of spherical c.m.\ and rel.\ coordinates all
 integrals could be reduced to products of one-dimensional
 integrals that can be solved very accurately and efficiently. In fact,
 the angular integrations are performed analytically. Moreover,
 the Gaussian quadrature provides exact results for all radial
 integrals except the ones of the interatomic interaction.  However,
 even the latter ones can be calculated to high precision using
 Gaussian quadrature. The compactness of the $B$ splines leads to
 sparseness of the Hamiltonian matrices, since only few integrals
 involve two non-zero $B$ splines. The compactness and thus the
 resulting band structure of the one-particle Hamiltonian matrices (of
 c.m.\ and rel.\ motion) is controlled by the order $k_{\rm rel.}$ and
 $k_{\rm c.m.}$ of the $B$ splines. A higher order leads to a broader
 band
 structure, but it offers also a higher flexibility of the basis
 functions, since it corresponds to a polynomial of higher order. Thus
 less $B$ splines are needed for a comparable result, if a higher order
 is adopted. Usually, the orders of $k_{\rm rel.}$ ($k_{\rm c.m.}$)
 of 8 or 9 turn out to be the best compromise with respect to basis-set
 size and sparseness, but also with respect to numerical stability in
 view of the finite precision in which floating-point numbers are
 stored in the computer.

 The two other parameters defining the radial $B$-spline basis 
 are the number of $B$ splines and their knot sequence. Clearly, 
 the computational efforts (number of integrals that have to 
 be calculated and size of the matrices that have to be diagonalized) 
 depend crucially on the number of $B$ splines. Since more $B$ splines 
 are required for describing highly oscillatory wavefunctions, 
 it is most efficient to use non-uniform knot sequences in which 
 the $B$-spline density is higher in the highly oscillatory regions 
 of (radial) space. 

 In the context of ultracold collisions the energetically 
 low-lying c.m.\ orbitals $\psi(\vec{R}\,)$ are the ones that usually 
 are of main interest. Since they possess only a small number of 
 nodes (none in the lowest state), the demands on the $B$-spline 
 basis are not too high. For the simplest case of a single-well 
 lattice potential $N_R=50$ to $N_R=100$ $B$ splines (or even less) were
 found to be  
 sufficient to obtain convergence for the ground and lowest-lying 
 excited states of the c.m.\ motion~\cite{cold:gris09}. Also the 
 lowest-lying states in a more structured trap potential like the 
 triple-well potential considered in \cite{cold:schn09} were 
 satisfactorily treated with 70 $B$ splines. A linear knot 
 sequence is usually adequate in this case. 

 The description of the rel.\ orbitals $\varphi(\vec{r}\,)$ is more demanding,
 since one is usually not interested in the lowest-lying, deeply bound
 molecular states, but in the most weakly bound states or the low-lying
 dissociative ones. The Born-Oppenheimer potentials of alkali-metal atom
 dimers support often a large number of bound states~\cite{cold:gris07}. The
 very long-ranged, weakly bound states consist, therefore, of a highly
 oscillatory inner part (covering the molecular regime and providing the
 orthogonality to all lower lying bound states) and a rather smooth long-range
 part.  Hence, it is advantageous to distribute a majority of $B$ splines in
 the molecular range of the potential while they are a sparser distributed in
 the outer part.  The distribution of the B splines is given by the knot
 sequence $\{r_i\}$, $i=1,2,\dots$ specifying a continuous chain of segments
 on which the $B$-spline functions are defined.
 The choice of a combined linear and geometrical knot
 sequence~\cite{bsp:vann04} for describing the short and long-range parts,
 respectively, has proven to be very efficient also in the present
 context~\cite{cold:gris09}.  The linear distribution of the knot sequence is
 given by
 \begin{equation}
  \ds r_{i+k_{\rm rel.}} = \rho_{\rm min.} + i\, s,\quad 
    i=1,2,...,N_r^{\rm lin.}-k_{\rm rel.}
    \label{eq:lin_ks}
  \end{equation}
 where $N_r^{\rm lin.}$ and $\rho_{\rm min.}$ are the number of $B$
 splines in the linear interval and the origin of the linear interval,
 respectively. Furthermore, $s$ is the linear step size, i.e., the distance between
 neighboring knots given as
 \begin{equation}
  \ds s = \frac{\rho_{\rm lin.}-\rho_{\rm min.}}
               {N_r^{\rm lin.}-k_{\rm rel.}+1}\,.
 \end{equation}
 Owing to the steep inner-wall of the molecular potential the wave
 function $\varphi_i(r)$ vanishes well before $r=0$. Therefore, 
 $\rho_{\rm min.}$ is usually taken non-zero in order to save on the
 number of $B$ splines. The exact value of $\rho_{\rm min.}$ depends on
 the potential of the considered electronic state. 
 Note, $k_r$ points
 must be placed at both ends of the box in accordance with
 the definition of $B$ splines on the knot sequence (e.g.,
 $r_1=r_{2}=...=r_{k_{\rm rel.}}=\rho_{\rm min.}$). The linear step $s$
 is taken as the scale factor for the geometric progression to ensure
 the smooth distribution of $B$ splines at the border of the linear and
 geometric zones.
 In the geometrical knot sequence the separation of the knot points increases
 (according to a geometric series) with increasing distance.  It is given by
 \begin{equation}
   \ds r_{i+N_r^{\rm lin.}} = \rho_{\rm lin.}+s\, q^{i-1},\quad
   i=1,...,N_r^{\rm geo.}
   \label{eq:geo_ks}
 \end{equation}
 where $N_r^{\rm geo.}$ is the number of $B$ splines in the
 geometric interval  and $q$ is the
 common ratio for the geometric progression defined as
 \begin{equation}
  \ds q = \left(
                \frac{\rho_{\rm box.}-\rho_{\rm lin.}}{s}
          \right)^{\ds \frac{1}{N_r^{\rm geo.}-1}}\,.
 \end{equation}

 The last parameter is the maximum value of angular momentum $l_{\rm max}$ 
 used and thus the number of angular basis functions. Clearly, a 
 more spherical-like lattice potential needs less angular momenta 
 for representing the orbitals. The worst case is a highly anisotropic 
 lattice geometry, since the spherical-harmonic basis converges 
 extremely slowly in such a case. While a small value of $l_{\rm max}$ 
 of 1 or 2 was sufficient to obtain converged orbitals for an
 isotropic 
 (cubic) single-well potential in \cite{cold:gris09} in which
 among others also the experiment in \cite{cold:ospe06a} was
 successfully modeled and thus realistic parameters were adopted. On
 the other hand, the much more anisotropic triple-well potential
 considered 
 in \cite{cold:schn09} required $l_{\rm max}=32$ for converged 
 orbitals.

\subsection{Example of a very anisotropic trap}
\label{subsec:example}

Since a highly anisotropic lattice geometry provides a great challenge to our
approach, it is important to demonstrate that even such a problem can be
handled satisfactorily. Another motivation is the present interest in
ultracold atomic systems of reduced dimensionality. Using strong confinement
along one or two orthogonal directions, quasi one- or two-dimensional
structures can nowadays be produced in the lab. Such systems show remarkable
quantum properties not encountered in three dimensions. A major experimental
breakthrough was the realization of Tonk-Girardeau gases of Bosons with
strongly repulsive interaction~\cite{cold:gira60,cold:pare04}. Being placed in
1D and repelling each other Bosons are hindered from occupying the same
position in space. This mimics the Pauli exclusion principle for Fermions,
causing the Bosonic particles to exhibit Fermionic properties. Another
peculiar property of reduced dimensionality systems is the occurrence of
confinement-induced resonances that were recently experimentally
observed~\cite{cold:hall10b,cold:froh11}.  Confinement-induced 1D Feshbach resonances
reachable by tuning the 1D coupling constant via 3D Feshbach scattering
resonances occur for both Bose gases~\cite{cold:olsh98} and spin-aligned Fermi
gases~\cite{cold:gran04}. Near a confinement-induced resonances, the effective
1D interaction is very strong, leading to strong short-range correlations,
breakdown of effective field theories, and emergence of a highly correlated
ground state. Although confinement-induced resonances were originally
predicted to occur already when only the rel.\ motion is
considered~\cite{cold:olsh98,cold:petr00,cold:berg03}, it was recently shown
and furthermore verified using the here presented numerical approach that the
ones observed in \cite{cold:hall10b} are in fact caused by the coupling of the
c.m.\ and rel.\ degrees of freedom~\cite{cold:sala11}.

 A typical system to study effects of low dimensionality is two atoms
 confined in a quasi 1D cigar-shaped harmonic trap. We adopt this
 two-body setup to probe the quality of the present computational
 method in an extreme case of the strong anisotropic confinement. As
 was already mentioned, if the interparticle interaction between atoms
 is modeled by the $\delta$-function pseudopotential the problem can be
 solved analytically. Therefore, the numerical results can be
 compared with the predictions of the pseudopotential model.
 One-dimensional external confinement is prepared by setting a strong
 harmonic frequency in either of two spatial directions, e.\,g.,
 $\omega_x=\omega_y \gg \omega_z$ where $\omega_i$ stands for the
 harmonic oscillator frequency along the spatial direction $i$. It is
 worthwhile to note that the analytical solution is only known for this
 special case of the single anisotropy~\cite{cold:lian08}.
 This emphasizes the need for numerical approaches to calculate spectra
 in totally anisotropic confinement where 
 $\omega_x \ne \omega_y \gg \omega_z$.

\begin{figure}[ht]
   \centering
   \includegraphics[width=0.46\textwidth]{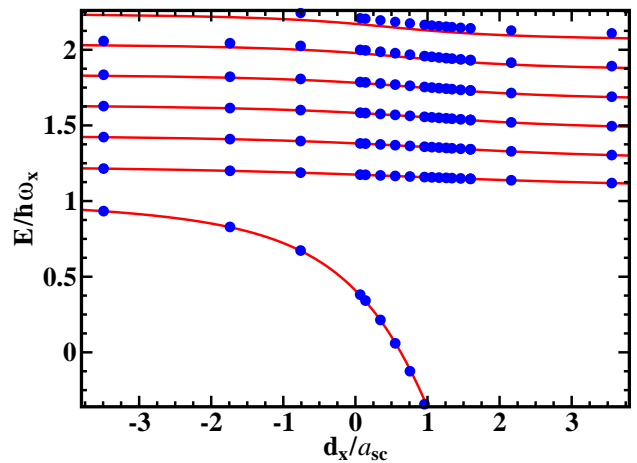}
   \caption{Energy spectrum of two Bosonic atoms confined in a
     harmonic trap
     of anisotropy ${\omega_x}/{\omega_z}=10$. The numerical
     calculation (blue dots) is compared to the analytical
     prediction of the pseudopotential approximation (red lines).}
   \label{fig:spectrum_solo}
 \end{figure}

 The Hamiltonian of two identical particles in our harmonic trap separates in
 rel.\ and c.m.\ motion. The c.m.\ spectrum reduces to the one of a simple
 harmonic oscillator. This simplifies the problem to calculate the rel.\
 spectrum. Hence, it is sufficient to solve numerically Eq.~(\ref{eq:schrel})
 only. In the following two Bosonic $^7$Li atoms are considered that are
 placed in the prolate trap with $\omega_x=\omega_y=10\,\omega_z$ and
 interacting via the potential of the $a^3 \Sigma_g^+$ electronic triplet
 state. This electronic state has the advantage of supporting a small number
 of bound states. Therefore, a smaller number of $B$ splines is required to
 reproduce the $\varphi_i(\vec{r})$ functions.  The numerical data for the
 Born-Oppenheimer potential curve are adopted from~\cite{cold:gris07}. In
 order to achieve converged results for the first few states $l_{\rm max}=30$,
 a box size of approximately $\rho_{\rm box}\approx 5\times 10^{4} a_0$, where
 $a_0$ is the Bohr radius, and $N_r=100$ $B$ splines of the order $k_{\rm
   rel.}=8$ were used. Half of the $B$ splines ($N_r^{\rm lin.}=50$) are
 distributed linearly according to Eq.~(\ref{eq:lin_ks}) over the small
 interatomic distance of $\rho_{\rm lin.}=15\,a_0$ to reproduce the highly
 oscillatory structure of $\varphi_i(r)$ in this region. 
 Furthermore, for our calculations we can safely choose 
 $\rho_{\rm min.}=2\,a_0$.
 The remaining $B$
 splines ($N_r^{\rm geo.}=50$) are distributed in an ascending geometric
 progression according to Eq.~(\ref{eq:geo_ks}) over the residual interval.
 Finally, different values of the interaction strength are obtained using a
 single-channel approach by a smooth variation of the inner wall of the
 molecular interaction potential~\cite{cold:gris10}.
 Figure~\ref{fig:spectrum_solo} shows the calculated spectrum of the rel.\
 Hamiltonian together with the analytical prediction of the pseudopotential
 approximation. The spectral curves are plotted as functions of $d_x/a_{\rm
   sc}$, where $d_x=\sqrt{\hbar/(\mu\, \omega_x)}$ is the harmonic oscillator
 length in transversal direction.
 As is evident from Fig.~\ref{fig:spectrum_solo}, the first four trap
 states and the bound state match perfectly with the analytical
 prediction of the pseudopotential model. However, high lying states
 show deviations.

\begin{figure}[ht]
   \centering
   \includegraphics[width=0.46\textwidth]{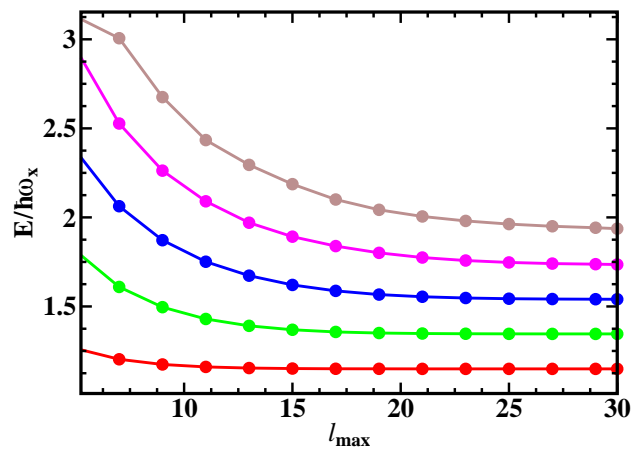}
   \caption{The energy of the first five
     trap states for different values of $l_{\rm max}$ in
     the strongly repulsive regime for
     $d_x/a_{\rm sc}=1.46$. The deeper lying bound state is not shown
     here, because it is already converged for $l_{\rm max}=5$.
     (To guide the eye, the discrete points of the numerical
     calculation  are connected by continuous lines.)
   \label{fig:spectrum_convergence}}
 \end{figure}

 In order to explain the increasing mismatch found for higher lying
 states and to give an impression of the convergence behavior of the
 present approach, energy spectra for different values of the
 angular momentum are calculated. For the calculations 
 $d_x/a_{\rm sc}=1.46$ is chosen. Close to this value a
 confinement-induced resonance should occur~\cite{cold:olsh98}
 motivating
 this choice.  The results of the calculations are shown in
 Fig.~\ref{fig:spectrum_convergence}. As is seen from this figure, at
 $l_{\rm max}=30$ the slopes of the energy curves are approximately
 zero,
 especially for the first four trap states. This indicates that
 convergence is achieved with respect to $l_{\rm max}$. Figure~\ref{fig:spectrum_convergence}
 clearly demonstrates
 that the high lying states require larger values of the angular
 momentum for
 convergence. This can now as well be concluded from
 Fig.~\ref{fig:spectrum_solo}.

 While convergence can be extended to higher energies or higher
 anisotropies by increasing $l_{\rm max}$, the computational efforts
 increase adding new angular momenta. However, as the precision is only
 limited by hardware capacities, this example demonstrates the
 applicability of the approach even for extreme setups.

\subsection{Exact diagonalization (configuration interaction)}
\label{subsec:exdiag}

 The demands of the exact diagonalization also known as configuration
 interaction (CI) depend evidently again on the physical system under
 consideration. While the atom-atom interaction is very efficiently
 handled even for strongly interacting atoms and basically contained in
 the rel.\ orbitals, the CI takes care of the c.m.\ and rel.\ coupling.
 Thus its convergence depends on the strength of this coupling. 
 Correspondingly, convergence is much faster for weakly coupled
 systems. If all configurations that can be built with the c.m.\ and
 rel.\ orbitals are included, full CI is achieved.  While being well
 defined, full CI scales very inconveniently with the number of
 orbitals. Therefore, it is in practice in most cases more advantageous
 to include only a limited number of the possible configurations. For
 example, the inclusion of the orbitals of very high energy 
 does usually not lead to a noticeable improvement of the low lying states.
 Therefore, an energy cut-off may be introduced in the orbital
 selection. If the interest is mainly on those states that are weakly
 bound or dissociative, but close to the dissociation threshold, the
 deeply bound molecular states can be omitted from the CI
 configurations~\cite{cold:gris09}. Note, however, that a good
 representation of those states in the calculation of the rel.\
 orbitals is nevertheless important, since otherwise the rel.\ orbitals
 of the weakly bound states are not well described and this can in
 practice not be compensated by the CI calculation. In the worst case
 the calculation of the rel.\ orbitals provides too few bound states
 and then the nodal structure of the weakly bound states is evidently
 wrong.

 Although the coupled Hamiltonian matrices are usually much larger than
 the uncoupled ones and are therefore harder to diagonalize, there are
 two points easing the treatment of the coupled ones. First, the
 eigenvalue problem to be solved in the CI step is a standard one
 (meaning that the the basis is orthonormal), while generalized ones
 occur in the orbital calculations. Second, in many cases only a
 relatively small number of CI states is required.  Therefore, it is
 possible to use Lanzcos or Davidson type diagonalization routines that
 (iteratively) provide a small number of eigenstates within a specific
 energy interval.  Here we adopted the Davidson-based diagonalization
 routine JADAMILU~\cite{gen:boll07} that is especially designed for the
 efficient diagonalization of large sparse matrices.

 Finally, it should be noted that the choice of expressing all 
 wavefunctions in rel.\ and c.m.\ coordinates is advantageous for 
 computational reasons, but often not very helpful in the 
 interpretation of the results. Especially in the case of 
 multi-well potentials the obtained wavefunctions and densities 
 are often very complicated. Therefore, a further code was 
 implemented that allows the application of the inverse 
 coordinate transform from c.m.-rel.\ coordinates to the absolute 
 ones according to $\ds\vec{r}_1  = \vec{R} + \mu_2 \vec{r}$ 
 and $\ds \vec{r}_2 = \vec{R} - \mu_1 \vec{r}$ to the wavefunctions, 
 especially to $\Psi(\vec{R},\vec{r})$ which allowed a much 
 easier physical interpretation, e.\,g., in \cite{cold:gris09}.

 \section{Summary and outlook}
 \label{sec:outlook}

 An approach that allows the full numerical description of two 
 ultracold atoms in a finite orthorhombic 3D optical lattice 
 is presented. The coupling between center-of-mass and relative
 motion coordinates is incorporated in a configuration-interaction 
 manner and hence the full 6D problem is solved. An important 
 feature is the use of realistic interatomic interaction potentials 
 adopting, e.\,g., numerically provided Born-Oppenheimer curves. 
 The use of spherical harmonics together with $B$ splines
 as basis functions and the expansion of the trap in terms of
 spherical harmonics leads to an analytical form of the matrix
 elements, except those of the interparticle interaction, if the 
 interaction potential is defined only numerically. The sparseness 
 of the Hamiltonian matrices due to the use of the compact radial 
 $B$-spline basis is considered explicitly. This makes the method 
 computationally efficient. Additionally, the lattice symmetry and 
 a possible indistinguishability of the atoms (Bosonic or Fermionic 
 statistics) is considered explicitly. This simplifies the calculations 
 and helps to interpret the solutions. 

 The here presented approach has already proven its applicability 
 by considering the influence of the anharmonicity in a single site 
 of an optical lattice in \cite{cold:gris09} where a corresponding 
 experiment \cite{cold:ospe06a} could be reproduced and analyzed. 
 The validity range of the Bose-Hubbard model together with an 
 improved determination of the Bose-Hubbard parameters was 
 investigated considering a triple-well potential in \cite{cold:schn09}. 
 
 The implemented approach was formulated in a rather general way 
 in order to allow extensions in various directions. Since the 
 optical-lattice potential is (via the Taylor expansion) expressed 
 as a superposition of polynomials, it is rather straightforward 
 to consider other than pure $\sin^2$ or $\cos^2$ potentials, as 
 long as they can be represented with (a reasonable number of) 
 polynomials. This includes tilted lattices and superlattices. 
 Care has, however, to be taken that the orthorhombic symmetry 
 is either preserved, or new symmetry rules have to be implemented.

 Substitution of the numerical interatomic potential by, e.\,g., 
 the Coulomb potential allows to describe either two electrons 
 or an exciton in quantum-dot atoms or molecules.
 It is furthermore planned to implement non-isotropic interatomic 
 potentials like dipole-dipole interactions as they are of interest 
 for Cr or Rydberg atoms and for heteronuclear diatomic molecules.    
 Finally, an extension of the approach in the direction of time-dependent 
 problems (with time dependent lattice or interatomic interaction 
 potentials) is currently under way.

\section*{Acknowledgments}
 The authors thank Prof.~Matthias Bollh\"ofer for providing the 
 code JADAMILU and helpful comments for its implementation and use. 
 The authors are grateful to the {\it Deutsche Forschungsgemeinschaft} 
 (within {\it Sonderforschungsbereich} SFB 450), the {\it Fonds der
 Chemischen Industrie}, and the {\it Humboldt Centre for Modern Optics} 
 for financial support. S. Grishkevich gratefully acknowledges the
 integrating project of the European Commission (AQUTE) for the
 finantial support.

% \bibliography{journals,cold,gen,vdw,bsp}

 \end{document}